%% file: mainInforms.tex
\documentclass[moor]{informs1}

\usepackage[colorlinks=true,breaklinks=true,bookmarks=true,urlcolor=blue,
     citecolor=blue,linkcolor=blue,bookmarksopen=false,draft=false]{hyperref}

\usepackage{natbib}
 \NatBibNumeric
 \def\bibfont{\small}%
 \bibpunct[, ]{[}{]}{,}{n}{}{,}%

\def\EMAIL#1{\href{mailto:#1}{#1}}

\usepackage[caption=false]{subfig}
\usepackage{amsbsy}

\DeclareMathOperator{\tr}{tr}
\DeclareMathOperator{\E}{\mathbb{E}}
\DeclareMathOperator{\M}{\mathbb{M}}
\DeclareMathOperator{\bigo}{\mathcal{O}}
\DeclareMathOperator{\diag}{diag}
\DeclareMathOperator{\as}{\quad\text{a.s.}}

\TheoremsNumberedThrough
{\theoremstyle{THkey}\newtheorem{theorem_nonum}{Theorem}}

\newboolean{showcomments}
\setboolean{showcomments}{true}

\newcommand{\babak}[1]{  \ifthenelse{\boolean{showcomments}}
{ \textcolor{red}{(Babak says:  #1)}} {}  }
\newcommand{\elizabeth}[1]{\ifthenelse{\boolean{showcomments}}
{ \textcolor{red}{(Elizabeth says: #1)} } {} }
\newcommand{\adam}[1]{\ifthenelse{\boolean{showcomments}}
{ \textcolor{red}{(Adam says:  #1)}}{}}
\newcommand{\bose}[1]{\ifthenelse{\boolean{showcomments}}
{ \textcolor{red}{(Bose says:  #1)}}{}}

\begin{document}
\TITLE{The Cost of an Epidemic over a Complex Network: \\A Random Matrix Approach\footnote{A preliminary version of this work was presented at GameNets, 2011 (\citet{gameNets11}).}}

\ARTICLEAUTHORS{%
\AUTHOR{Subhonmesh Bose}
\AFF{Dept. of Electrical Engineering, California Institute of Technology, \EMAIL{boses@caltech.edu}}
\AUTHOR{Elizabeth Bodine-Baron}
\AFF{RAND Corporation, \EMAIL{Elizabeth\_Bodine-Baron@rand.org}}
\AUTHOR{Babak Hassibi}
\AFF{Dept. of Electrical Engineering, California Institute of Technology, \EMAIL{hassibi@systems.caltech.edu}}
\AUTHOR{Adam Wierman}
\AFF{Dept. of Computing and Mathematical Sciences, California Institute of Technology, \EMAIL{adamw@caltech.edu}}
} 

\ABSTRACT{%
In this paper we quantify the total economic impact of an epidemic over a complex network using tools from random matrix theory.  Incorporating the direct and indirect costs of infection, we calculate the disease cost in the large graph limit for an SIS (Susceptible -- Infected -- Susceptible) infection process.  We also give an upper bound on this cost for arbitrary finite graphs and illustrate both calculated costs using extensive simulations on random and real-world networks.  We extend these calculations by considering the total social cost of an epidemic, accounting for both the immunization and disease costs for various immunization strategies and determining the optimal immunization.  Our work focuses on the transient behavior of the epidemic, in contrast to previous research, which typically focuses on determining the steady-state system equilibrium.

}%



\maketitle

\input{intro}
\input{netmodel}
\input{infmodel}

\input{results}
\input{sim_disc}
\input{extensions}
\input{conclusion}

\begin{APPENDIX}{Proof of Lemma \ref{lemma:allTerms}}
\input{appProofs}
\end{APPENDIX}

\section*{Acknowledgments.}
The authors would like to thank Professor K. Mani Chandy and Prof. Leeat Yariv for useful comments and suggestions.  This work was supported in part by the National Science Foundation under grants NSF CNS-0846025, CCF-0729203, CNS-0932428 and CCF-1018927, by the Office of Naval Research under the MURI grant N00014-08-1-0747, and by Caltech's Lee Center for Advanced Networking.

\bibliography{ref}

\end{document}

%% file: intro.tex
\section{Introduction}

Epidemic models attempting to quantify how diseases are transmitted have been extensively studied since the SIR (Susceptible-Infectious-Recovered) model was proposed in 1927 in \citet{Kermack27}.  Though initially these models were proposed to understand the spread of contagious diseases, the insights learned from them apply to many other settings where something spreads through a population of agents.  For example, applications such as (i)  network security, where the goal is to understand and limit the spread of computer viruses, e.g., in \citet{Chen05}, \citet{WangKnight00}, \citet{Cohen03}  (ii) viral advertising, where the goal is to create an epidemic to propagate interest in a product, e.g., in \citet{Phelps04}, \citet{Richardson02}, and (iii) information propagation, where the goal is to understand how quickly new ideas propagate through a network, e.g., in \citet{Huang06}, \citet{Jacquet10}, \citet{Cha09}, can all be understood through the lens of epidemic models.  See \citet{Jackson08} and \citet{Daley05} for comprehensive surveys of prior results.

Most epidemic models focus on determining the existence and stability of system equilibria for various diseases, applying  Lyapunov's stability theory to the SIS (Susceptible-Infected-Susceptible) infection model.  Early models assume a well-mixed population (\citet{AndersonMay91}), i.e., any node can infect any other node.   In practice, however, this is rarely the case, motivating the study of epidemics where the interaction of the agents is limited to a network, such as \citet{Vesp02b, Pastor01, Vesp02, Vesp03, Vesp04, wang03, chak08}.  Some of this work examines possible containment or immunization schemes to minimize the final number of infected nodes, or eradicate the disease entirely.  See \citet{chak08, peng10, Chen05, Miller07} for some of these results.  Other work applies techniques from percolation theory to the SIR model, attempting to answer  whether an infection can start from a random node in a network and infect a giant component of the graph, e.g., \citet{moore2000epidemics, newman2002spread, kenah07pre}. Regardless of the infection model, most of this work focusses on the long-term behavior of the system.

Though understanding the extent to which an infection spreads is an important question in itself, these models can be even more useful in understanding the \emph{cost} of an epidemic. Within the medical community, there is a growing trend to quantify the cost of an epidemic by looking at the direct and indirect medical costs to both the hospitals and doctors treating and immunizing a population for specific diseases, as well as the cost to individuals in the population paying for medical care.   See \citet{Bownds03} and \citet{Rubin99} for two examples of such studies.  This interest in cost, both the cost of disease and the cost of immunization, is the motivation for the current paper.  

There is little existing work in the modeling community studying this cost, since any such calculation depends on the \emph{transient behavior} of the epidemic model that is often hard to analyze mathematically. This paper attempts to fill this void. Here, we assume an SIS model of infection, as in \citet{wang03}, \citet{chak08}, and \citet{peng10}, on a random network that is a variant of the generalized Erd\"{o}s-R\'{e}nyi random graph with arbitrary degree distributions as in \citet{chungBook, NewmanChapter} and define the cost or the economic impact of such an epidemic. Our main contribution is the derivation of (i) the exact cost of an epidemic in the large graph limit (Theorem \ref{thm:general}) and (ii) bounds on this cost for a given graph (Theorem \ref{thm:bound}). We further provide an optimal scheme for random one-time vaccination, minimizing the total cost of the epidemic, including both disease and immunization costs. All our results are validated via simulations.

A key feature of this paper is the technical approach used in the derivation of our results.  In particular, one of the initial motivations for our study was to highlight that tools from \emph{random matrix theory} can be adapted to provide powerful tools for the study of epidemics.  Random matrix theory is, by now, a rich area of mathematics with broad applications.   In particular, random matrix theory has found applications in wireless communications (\citet{tulino2004random}) and in the analysis of random graphs (\citet{van2011graph}).  We refer the reader to \citet{edelman2005random} and \citet{anderson2010introduction} for further details on this subject. In this paper, we show that tools from random matrix theory can be adapted to the study of epidemics.  In particular, we apply ideas from the Stieltjes transform in \citet{tulino2004random, taoBook} to analyze the transient behavior of an epidemic process over a random network. To the best of our knowledge, our approach using random matrices is novel to the study of epidemic processes.  Our study highlights that random matrix theory can provide powerful tools for this area, but they require significant adaptation to be applied.  Thus, we hope that this paper spurs future research that continues to develop this connection.

The paper is organized as follows. We introduce a random network model in Section \ref{sec:netmodel} and the infection process in Section \ref{sec:infmodel}. Using this framework, we define and compute the cost of an epidemic in Section \ref{sec:results}. In Section  \ref{sec:sim_disc},  we verify the assumptions of our model and results through extensive simulations. Finally, we discuss extensions and conclude in Sections \ref{sec:extensions} and \ref{sec:conc}, respectively. 

%% file: netmodel.tex
\section{Network Model}
\label{sec:netmodel}

There are two major components to the model studied in this paper: the model of the underlying network and the model of the infection process.  We discuss the model of the network here and then move to the model of the infection process in the next section.

Our network model is related to the ``configuration'' model in \citet{NewmanChapter} and the ``general random graph'' model from \citet{chungBook}; however, the model used is slightly more general than each. In particular, let $A$ be an $n \times n$ adjacency matrix corresponding to the network, where there are $n$ nodes in the population and $A_{ij} = 1$ if there exists a relationship from node $i$ to node $j$.  For the purposes of this paper, we only consider undirected graphs; i.e., $A_{ij} = A_{ji}$.  We assume that the network is drawn from a general class of random graphs, e.g., the network represented by $A$ could be a realization of an Erd\"{o}s-R\'{e}nyi random graph, $G_{n,p}$, which would correspond to allowing each edge to exist independently with probability $p$.

The construction of the graph proceeds as follows. First, define a degree distribution $p_n(\cdot)$, and obtain $n$ i.i.d.\ samples $w=(w_1,\ldots,w_n)$.  From this vector, generate a random graph given by the adjacency matrix:
\begin{equation}
A_{ij} = A_{ji} = \begin{cases} 1 & \text{ w.p. } w_i w_j \rho, \\
       0 & \text{ w.p. } 1-w_i w_j \rho,
       \end{cases} \text{ where } \rho = \frac{1}{\sum_i w_i}.
\label{eq:randgraph}
\end{equation}
Note that the expected degree of node $i$ is $\sum_j w_i w_j \rho = w_i$.  Since this model is fully determined by the degree distribution $p_n(\cdot)$, for ease of reference, we call it $G_{n,p_n(\cdot)}$.

\begin{figure}[htb] \centering
 \subfloat[$n=100, p= 0.0561$]{\label{fig:er100}\includegraphics[width=0.5\textwidth]{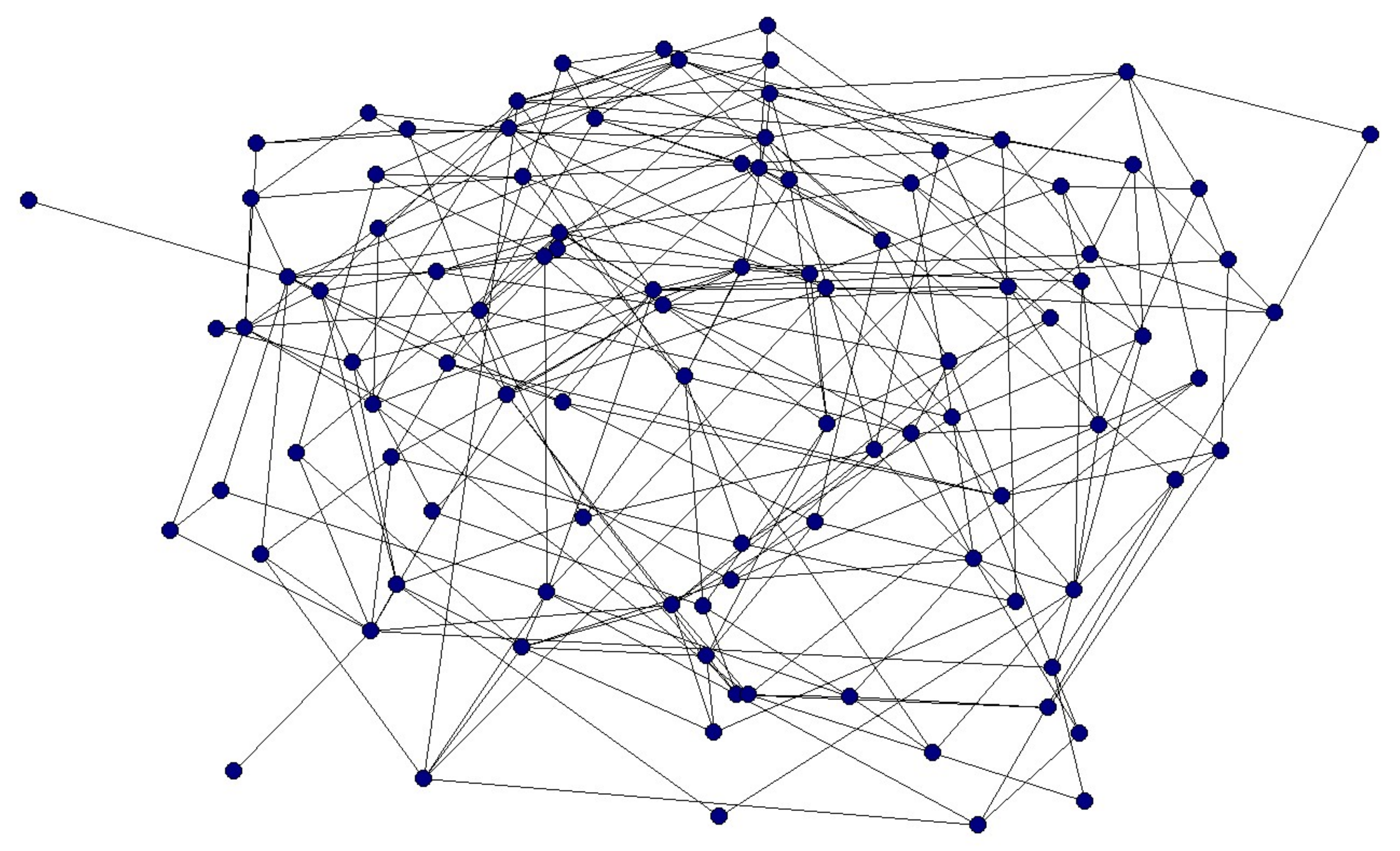}}
 \subfloat[$n=1000, p = 0.0169$]{\label{fig:er1000}\includegraphics[width=0.5\textwidth]{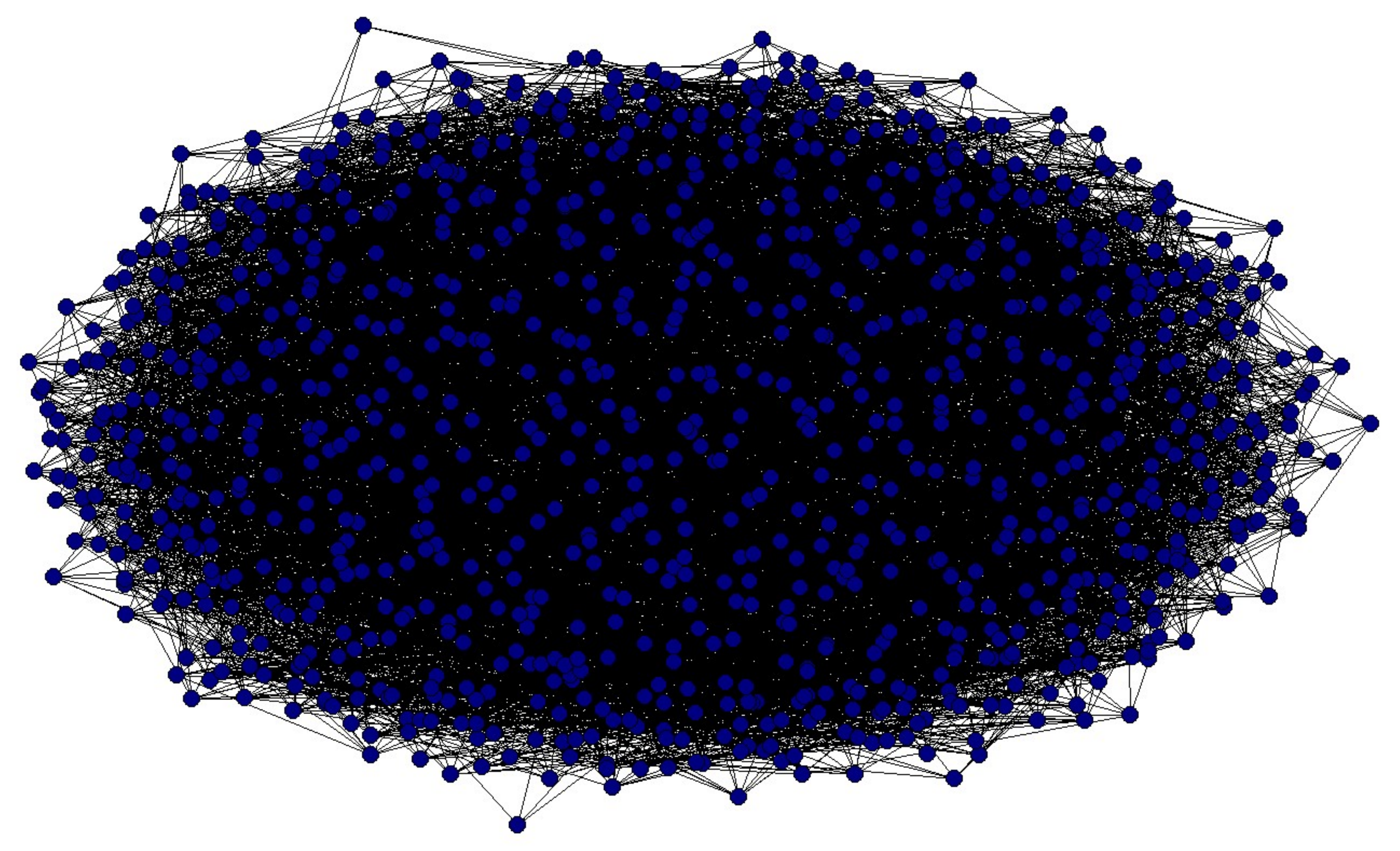}}
  \caption{Sample Erd\"{o}s-R\'{e}nyi random graphs}
  \label{fig:ER_gen}
\end{figure}

\begin{example} \label{ex:ER}
To generate the Erd\"{o}s-R\'{e}nyi random graph $G_{n,p}$, let $p_n(w) = \boldsymbol{\delta}(w-np)$, where $\boldsymbol{\delta}(\cdot)$ is the Dirac $\delta$-function.  Thus, $w = (np, np, \dots, np)$ and for all nodes $i$ and $j$, $A_{ij} = 1$ with probability $p$.  With our notation, we denote the graph as $G_{n, \boldsymbol{\delta}(w-np)}$.  Two example networks generated according to our model are shown in Figure \ref{fig:ER_gen}; the parameter $p$ is chosen just beyond the threshold for connectivity.
\end{example}

\begin{figure}[htb] \centering
 \subfloat[$n=100$]{\label{fig:exp100}\includegraphics[width=0.5\textwidth]{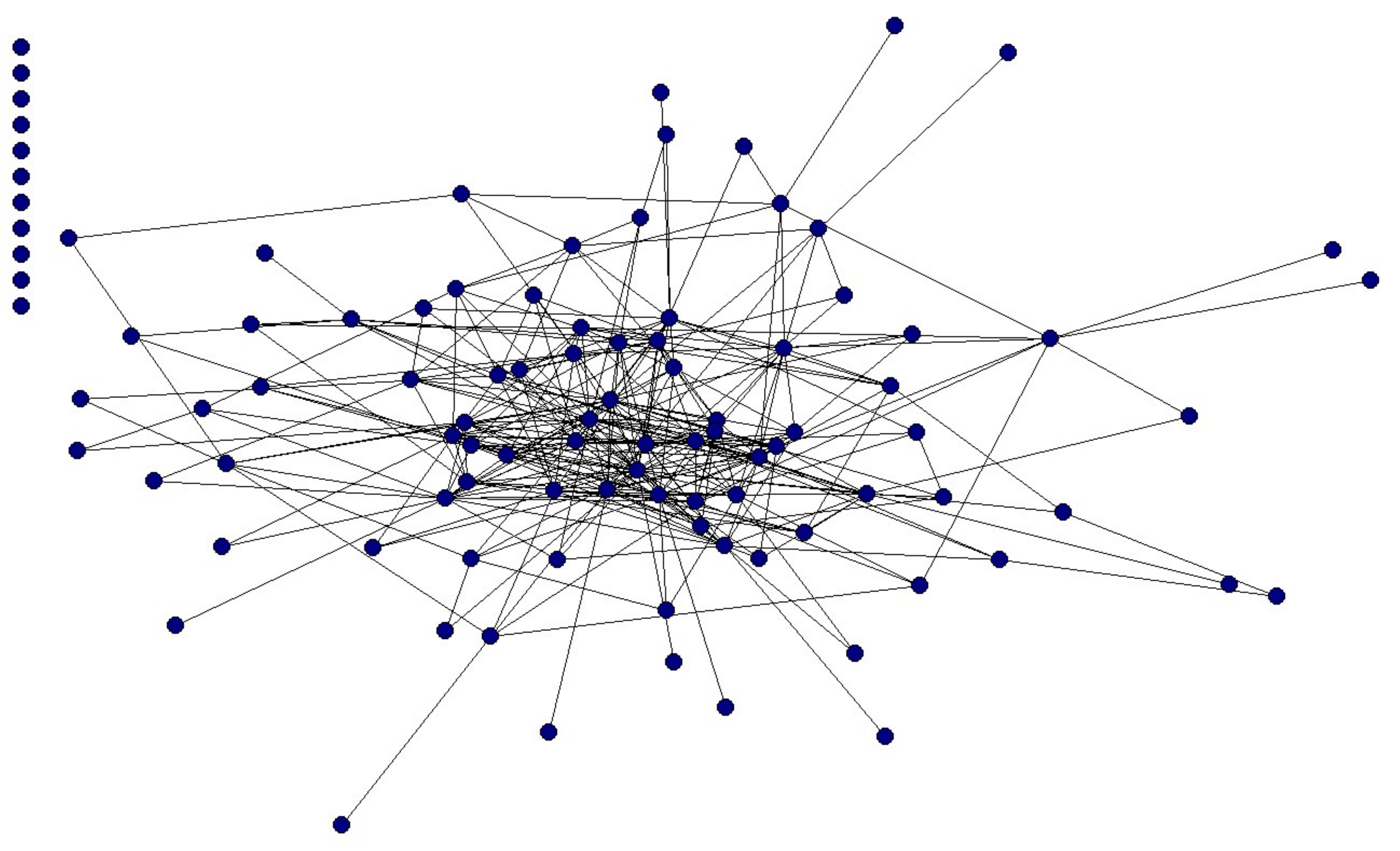}}
 \subfloat[$n=1000$]{\label{fig:exp1000}\includegraphics[width=0.5\textwidth]{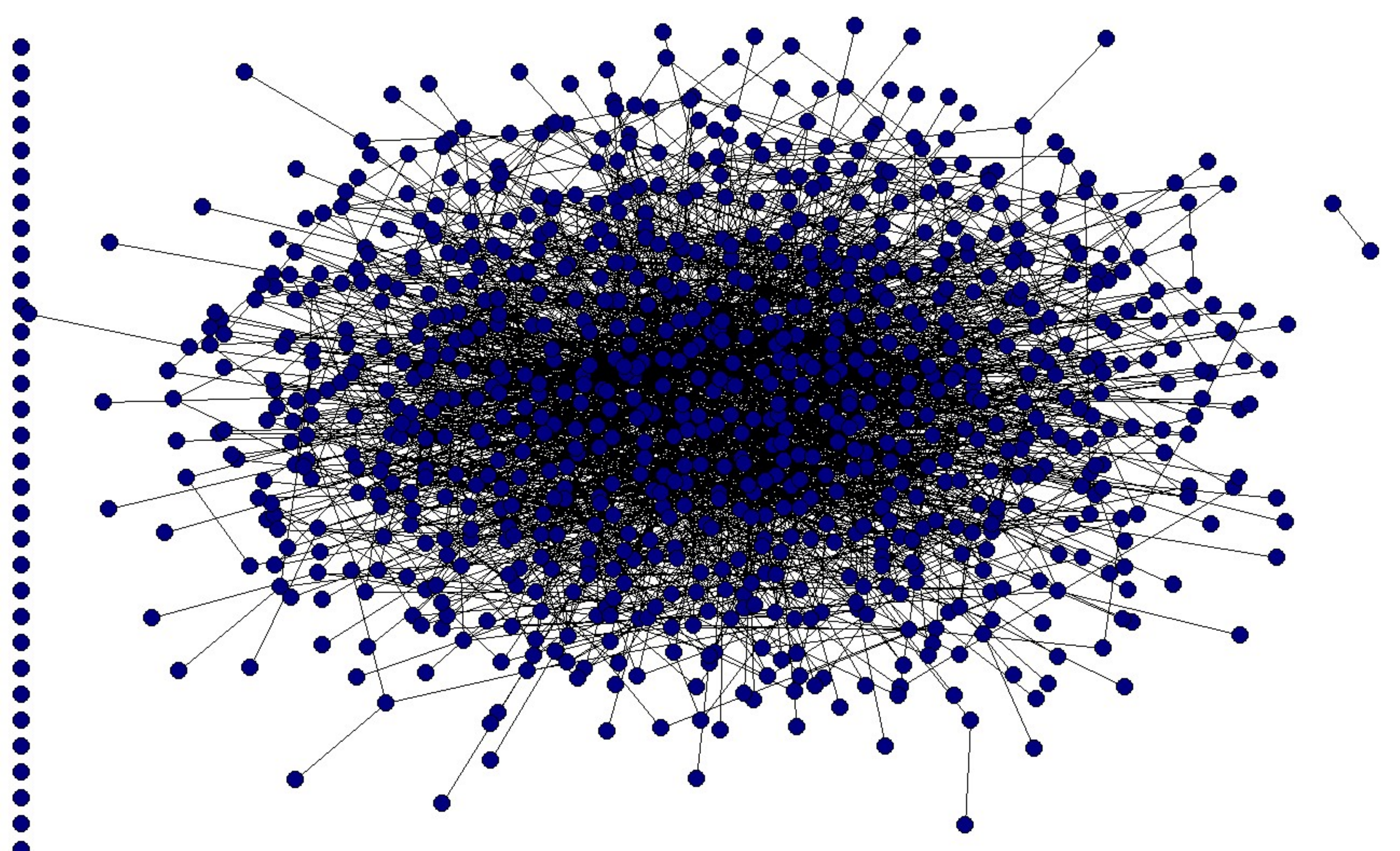}}
  \caption{Sample Exponential random graphs with $\lambda = \frac{1}{6}$}
  \label{fig:Exp_gen}
\end{figure}

\begin{example} \label{ex:Exp}
To generate a random graph with an exponential degree distribution, let $p(w) = \lambda e^{-\lambda w}$.  Following the construction outlined above, the resulting graph will have $n$ nodes with average degree $\lambda^{-1}$.  Example graphs with $100$ and $1000$ nodes and mean degree $6$ ($\lambda = 1/6$) are shown in Figure \ref{fig:Exp_gen}.
\end{example}

\begin{figure}[htb] \centering
 \subfloat[$n=100$]{\label{fig:pl100}\includegraphics[width=0.5\textwidth]{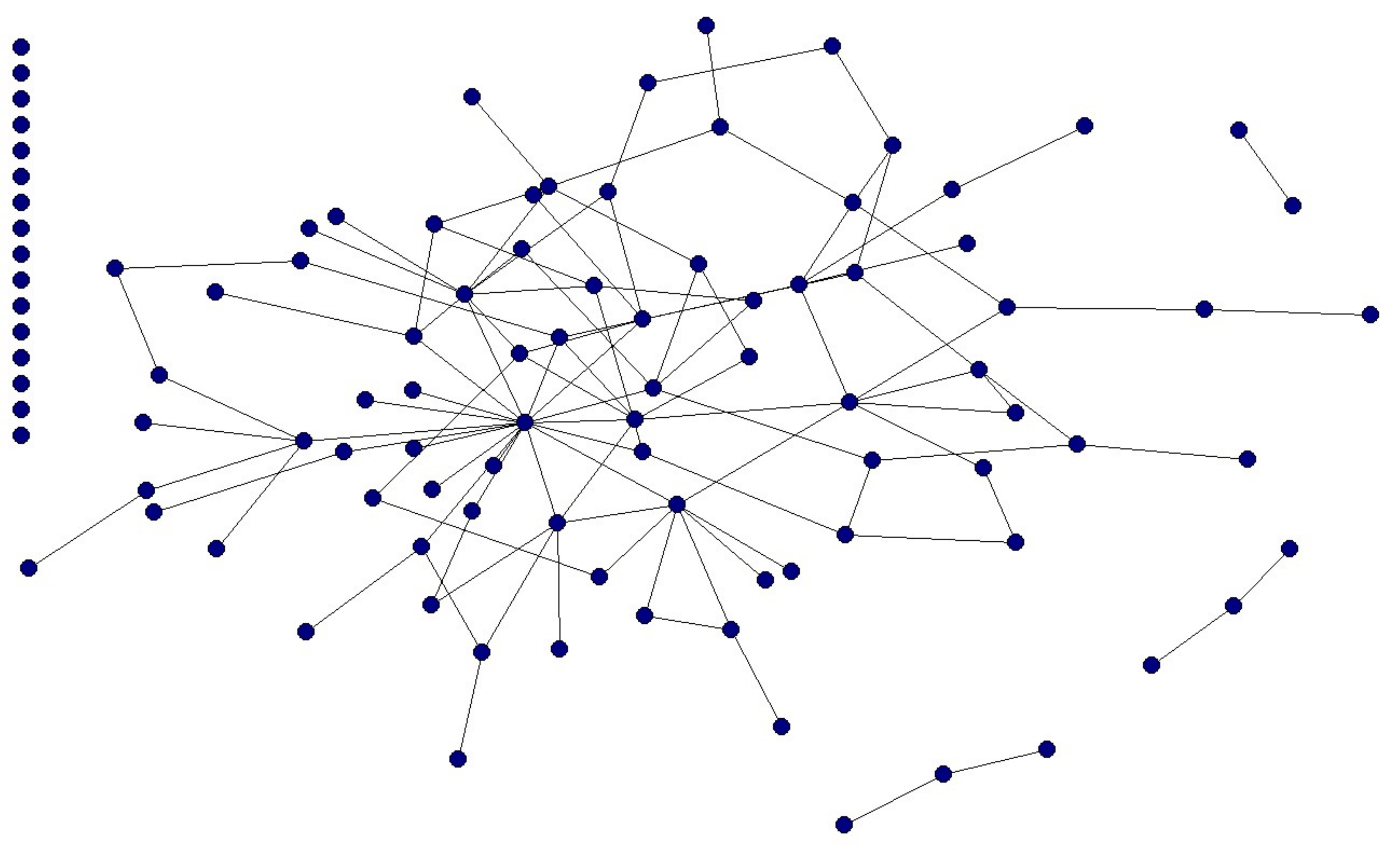}}
 \subfloat[$n=1000$]{\label{fig:pl1000}\includegraphics[width=0.5\textwidth]{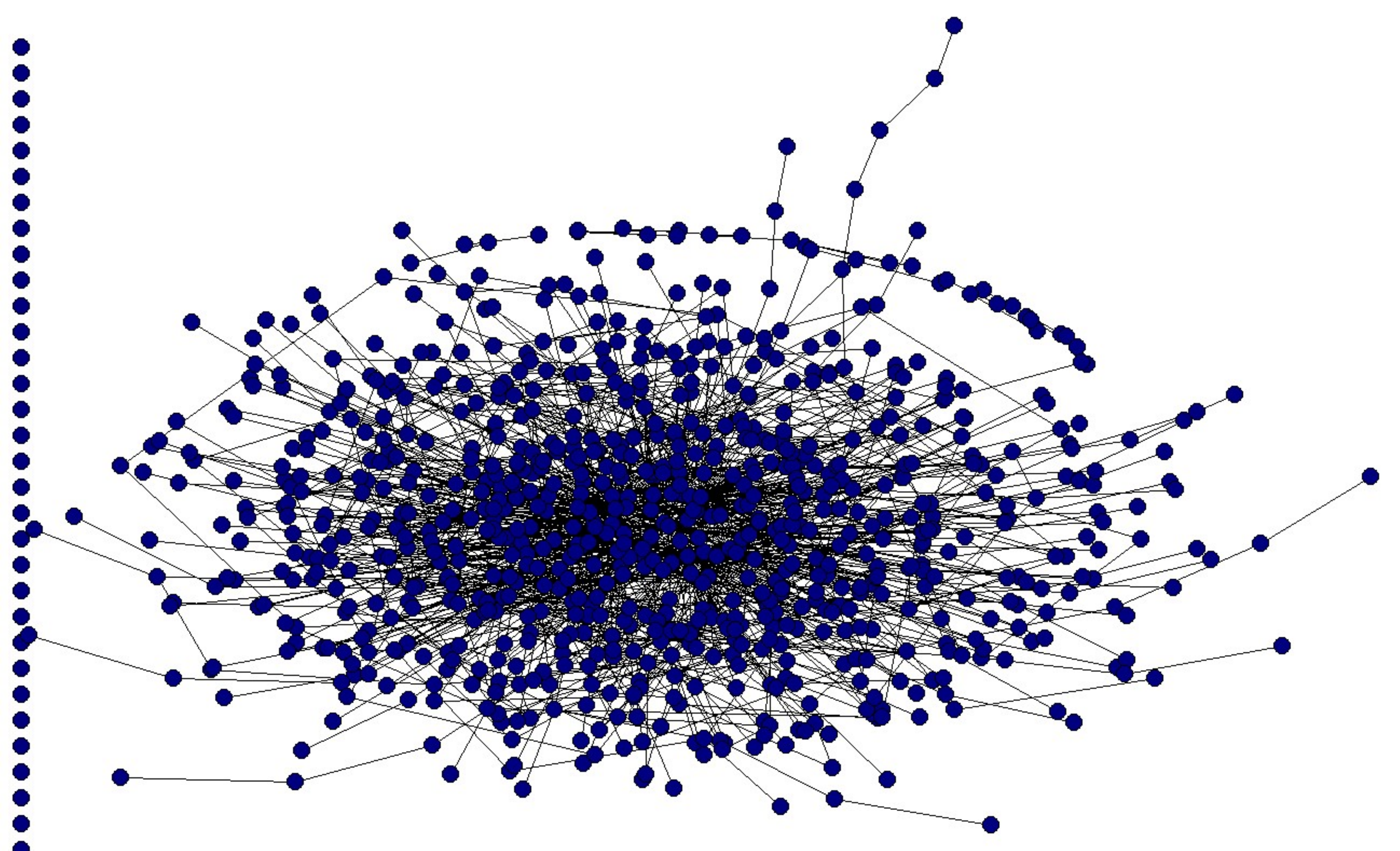}}
  \caption{Sample power law (Pareto, $\theta = 1.5$) random graphs}
  \label{fig:PL_gen}
\end{figure}

\begin{example} \label{ex:PL}
To generate a random graph with a power law degree distribution, (specifically, a Pareto distribution), let $p(w) = \frac{\theta}{w^{\theta+1}}$.  Following the construction outlined above, the resulting graph will have $n$ nodes with average degree $\frac{\theta}{\theta -1}$.  Two example graphs with $\theta = 1.5$ (mean degree $3$) are shown in Figure \ref{fig:PL_gen}. \end{example}

The network model $G_{n,p_n(\cdot)}$ is quite general.  To relate this model to the configuration model in \citet{NewmanChapter} and the general random graph model in \citet{chungBook}, note that in the case of the configuration model the degree sequence is enforced deterministically and that in the general random graph model the expected degree sequence is fixed rather than being drawn from a distribution.  However, note (like the two models it generalizes) the model we consider here does not exhibit clustering.  

%% file: infmodel.tex
\section{Infection Model}
\label{sec:infmodel}

In this section, we describe the infection process that is the focus of this paper.  It is based on the SIS (Susceptible-Infected-Susceptible) epidemic model. In this formulation, each node in the population transitions between two possible states, i.e., \emph{susceptible} and \emph{infected}. The process is characterized by two parameters, i.e., $\delta$ and $\beta$ that represent the recovery rate and the infection rate, respectively. Time is taken to be discrete and events proceed in each time-step as follows:
\begin{enumerate}
\item If node $i$ is infected, it recovers with probability $\delta$. Note that it cannot be infected in the same time-step in which it recovers.
\item If node $i$ is susceptible, it becomes infected by each of its neighbors with i.i.d.\ probability $\beta$.
\end{enumerate}

In its full generality, this model of SIS infection spread on a network is difficult to analyze. Thus, it is standard to make simplifying approximations to arrive at a mathematically tractable formulation. To that end, we adopt the so-called ``linear approximation'' commonly used in the literature, first derived in \citet{wang03}. 

To introduce the model, we start with some notation. Consider a graph $G$ on $n$ nodes over which the epidemic process runs. Let $A_{n \times n}$ denote its adjacency matrix. Note that the graph can be one sample of the random graph model presented in the previous section or it might just be a fixed graph. Define the $n \times 1$ vector $P(t)$ where $P_i (t)$ denotes the probability that node $i$ is infected at time $t$. Using this notation, the linear dynamics of the infection process are given as:
\begin{align}
P(t+1) & = \left[ \underbrace{(1-\delta)I }_{:= M_1}+ \underbrace{\beta A}_{:= M_2} \right] P(t). \label{eq:linModel}
\end{align}
where $I$ is the $n \times n$ identity matrix.  The probability of infection at time $t+1$ has contributions from two terms, i.e., $M_1 P(t)$ and $M_2 P(t)$. The first term is the contribution from the nodes that are infected at time $t$ and do not recover in the next time step with probability $1-\delta$. Infected neighbors contributes to the second term $M_2 P(t)$ through the adjacency matrix of the graph. Define the \emph{system matrix} of the epidemic process as:
\begin{align}
M = (1-\delta) I + \beta A. \label{eq:systemmat}
\end{align}
Note that in the special case when the infection begins with $\alpha$-fraction of the nodes infected at time $t=0$, we have  $P(0) = \alpha 1$, where `$1$' denotes an $n \times 1$ vector of all ones. Thus, $P(t)$ can be re-written as:
\begin{align} P(t) = M^t P(0) = \alpha M^t 1 . \end{align}
\noindent The above linear system is a commonly adopted approximation for the SIS model and, has been used in the literature, e.g., in \citet{chak08,peng10,wang03}.

We remark that the infection process model is a discrete-time one, where many authors have considered continuous-time models, e.g., \citet{Vesp02, Vesp02b, Vesp03, Vesp04}.  Interestingly, the mean field equation describing the system dynamics in the continuous time setting  is an exact analog of \eqref{eq:approx} and hence the results of our analysis can be generalized to the continuous-time case.

We now provide a derivation of \eqref{eq:linModel} that clearly delineates all the approximations involved in going from the networked SIS model to the linear system model in \eqref{eq:linModel}.  Let ${\cal N}_i$ be the set of neighbors of the $i$-th node. Consider a sample path $s$ of the disease propagation. In the sample path $s$, let $P_i ^{(s)}(t)$ denote the probability of node $i$ being infected at time $t$ in $s$. We analyze the quantity $P_i ^{(s)}(t+1)$ by conditioning it on the state of the node $i$ at time $t$ in $s$. Define the random variable ${X}_i^{(s)} (t)$ as the number of infected neighbors of $i$ if it is not infected at $t$ in $s$. If node $i$ is infected at $t$ in $s$, we set ${X}_i^{(s)} (t) = 0$. Thus, the probability of node $i$ getting infected at time $t+1$ given that it was not infected at time $t$ is $(1 - \beta)^{{X}_i^{(s)} (t)}$. Provided $\beta$ is small, this can be approximated by $1 - \beta  {X}_i^{(s)} (t)$. In summary, we can write an expression for $P_i ^{(s)}(t+1)$ as:
\begin{align}
P_i ^{(s)}(t+1) = (1-\delta)P_i^{(s)}(t)+ \beta  {X}_i^{(s)} (t)  (1-P_i^{(s)}(t)). \label{eq:samplePath}
\end{align}
\noindent We take expectation over all sample paths in \eqref{eq:samplePath} and make use of the fact that $P_i (t)$ is the expectation of $P_i ^{(s)}(t)$ over all sample paths to obtain:
\begin{align}
P_i (t+1) =  (1-\delta)P_i (t) + \beta \E_{s} \left[ \left(1 - P_i^{(s)} (t) \right) \cdot  {X}_i^{(s)} (t)  \right] \label{eq:noapprox}
\end{align}
\noindent Now, we approximate by assuming the terms inside the expectation in \eqref{eq:noapprox} to be independent:
\begin{align}
P_i (t+1) \approx & (1-\delta)P_i (t) + \beta (1 - P_i (t)) \cdot \E_s [ { X}_i (t) ].
\end{align}
The terms inside the expectation in \eqref{eq:noapprox} are not generally independent; however, such an approximation is ``reasonable'' when ${ X}_i^{(s)} (t)$ concentrates around its mean over all sample paths.  One may expect this to be true for graphs with light-tailed degree distributions, but not for heavy-tailed degree distributions.

Next, we evaluate $\E_s [ { X}_i^{(s)} (t)  ] $. Note that
it is the sum of the probabilities of the neighbors of node $i$ being infected given that node $i$ itself is susceptible at time $t$ in $s$. We approximate this quantity by dropping the conditioning on node $i$ being susceptible at $t$ in $s$.
This approximation is again valid if the behavior of this sample path is not too different from the evolution of the mean of all sample paths. Again, in a graph with light-tailed degree distribution, we expect this to be a ``reasonable'' assumption because no single event like a central node getting infected can cause large deviations in sample path behavior. Thus, we have $\E_s [ { X}_i^{(s)} (t)  ]  \approx \sum_{j\in{\cal N}_i} P_j (t)$. Combining this, we get the following non-linear recursion:
\begin{align}
P_i (t+1) \approx (1-\delta)P_i(t) + \beta (1-P_i(t)) \sum_{j\in{\cal N}_i}P_j(t) \label{eq:approx}
\end{align}
The above relation is a non-linear dynamical model. The quantity $1 - P_i(t) \leq 1$ and hence
\begin{align}
P_i (t+1) \leq (1-\delta)P_i(t) + \beta \sum_{j\in{\cal N}_i}P_j(t) \label{eq:approx.2}
\end{align}
We analyze the behavior of this linearized system that provides an upper bound to the evolution of the actual system. Expressed in matrix-vector form, this linearized dynamical model we study is given by the relation in \eqref{eq:linModel}. From \citet{lyapunov1992general}, the stability of the linearized model in \eqref{eq:linModel} around the vector of all zeros is sufficient to guarantee the stability of the non-linear system around the same vector, i.e., the disease dies out in the non-linear model \emph{iff} it dies out in the linear model. Thus, it is ``reasonable'' to use the linearized approximate model for the scenarios where the disease dies out. We relate this fact to the eigenvalues of the system matrix $M$ below (as in \citet{chak08,peng10,wang03}) and use it to analyze the infection process.

In summary, we use the above ``linear approximation'' for the SIS infection model in the following.  This is a natural, and standard, approximation.  However, as we have described, we should not expect the linear system in \eqref{eq:linModel} to accurately model SIS infection spread in all settings.  In particular, we should expect it to be a good approximation when the infection rate is small and when the degree distribution is light-tailed, but we should expect the accuracy to degrade as the infection rate grows or the tail of the degree distribution becomes heavier.  In Section \ref{sec:sim_disc}, we provide simulations to better understand the relationship of the accuracy of approximation with these parameters of the epidemic model.



%% file: results.tex
\section{Epidemic cost}
\label{sec:results}

Given the network and infection models described previously, we can now discuss the \emph{cost} of an epidemic on a network. As mentioned previously, a key contribution of this paper is to provide analytic results characterizing the cost of an epidemic over its entire lifetime.  This includes the effects of the \emph{transient} behavior of the epidemic, which is typically difficult to study.

To determine the cost of a disease, we consider a simple model where $c_d$ is defined as the cost of an individual being infected during a single time-step.  Thus, $c_d$ can capture both the direct costs to the individual for medication, doctor visits, etc., as well as secondary costs such as missed work.  We note that this model leaves open the question of how exactly to determine the parameter $c_d$.  This is ongoing work within the medical community; see \citet{Bownds03, Rubin99} for studies in this area.  In future work we expect to incorporate these results to obtain a more accurate cost of various diseases, but for the purposes of this paper, we leave it as a general parameter of the model.  Note also that this section only concerns the cost of the disease; it does not include the cost of any strategy to contain or control the epidemic, such as immunization.  In Section \ref{sec:extensions} we discuss minimizing the total cost of both the disease and the given containment strategy.

Given this model for the cost of disease to an individual, we can formalize the total social cost of an epidemic. To begin, assume some fraction $\alpha < 1$ of the nodes are infected at time $t=0$.  Denote the epidemic process on a network by the 5-tuple $\left(G, \delta, \beta, \alpha, c_d \right)$, where $G$ is the network, $\delta, \beta$ and $\alpha$ define the infection parameters, and $c_d$ defines the cost parameter. Define $C_D(n)$, the ``disease cost,''  as the expected (averaged over the random spread of the disease) per node disease cost of an epidemic \emph{during its entire course}.  Since the infection propagation is stochastic in nature, the cost for a given tuple $\left(G, \delta, \beta, \alpha, c_d \right)$ will be a random variable, and $C_D (n)$ denotes the expected value of this quantity when averaged over all infection propagation paths. To express it in closed form, note that the expected per node disease cost in a \emph{given time-step t} is simply $\frac{1^T P(t)}{n}$.  Furthermore, since $P(0) = \alpha {1}$ and $P(t) = M^{t-1} P(0)$, we can express the disease cost per node as
\begin{equation} C_D(n) := \frac{1}{n} \left[ {1}^T \left(\sum_{t=0}^\infty M^t \right) {1} \alpha c_d \right] . \end{equation}
The infinite sum converges if and only if the spectral norm of $M$ (maximum absolute eigenvalue of $M$) is less than 1 (\citet{horn2005matrix}). In that case, the disease will eventually die out, and we have
\begin{equation} C_D(n) = \frac{1}{n} \alpha c_d [{1}^T (I -M)^{-1} {1}]. \label{eq:diseaseCost} \end{equation}
We emphasize that the above expression is averaged over all possible infection propagation paths, but is a random variable when the underlying network is a random graph. However, we show that this cost converges almost surely to a deterministic constant when the network is drawn randomly according to our model (under certain conditions) and can be explicitly computed.

In particular, in Section \ref{subsec:asymptotics} we explore the cost in \eqref{eq:diseaseCost} in the asymptotic regime, i.e., in the large graph limit as $n \to \infty$, letting the degree distribution and the infection rate to vary with the population size. Specifically, the degree distribution for a population of size $n$ is $p_n(\cdot)$ and the infection rate is $\beta_n$. We compute $C_D (n)$ associated with the epidemic process defined by $\left( G_{n, p_n(\cdot)}, \delta, \beta_n, \alpha, c_d \right)$. Note that a fixed $n$ is a special case; i.e., the case where the degree distribution and infection rate do not scale with $n$ is subsumed in our result.  Due to the complexity of the asymptotic results, in Section \ref{subsec:boundFixedGraph} we provide a bound for the cost of the disease over a fixed graph. Further, in Section \ref{sec:sim_disc}, we illustrate these results through extensive simulations.

\subsection{Asymptotic cost of disease over random graph}
\label{subsec:asymptotics}

In this section, we compute the cost of the epidemic process $\left( G_{n, p_n(\cdot)}, \delta, \beta_n, \alpha, c_d \right)$, presenting our result formally in Theorem \ref{thm:general}.

Let $w_{n \times 1}$ be $n$ independent samples drawn according to the degree distribution $p_n(\cdot)$ and $W = \diag (w)$. Consider the vector $v := \beta_n w$. Assume that $p_n(\cdot)$ and $\beta_n$ scale such that the vector $v$ behaves as $n$ independent samples drawn from a scale invariant distribution $p(\cdot)$ that has a bounded support $[v_{min}, v_{max}]$, where $v_{max} > v_{min} > 0$. As explained below, this does not mean that $p_n(\cdot)$ is bounded, but when scaled with $\beta_n$, the limiting distribution has a bounded support.

Define the following quantities:
\begin{align}
V & := \diag(v) = \beta_n W, \label{eq:defV}\\
\mu & := \left(\displaystyle\sum_{i=1}^n v_i \right)^{-1}, \notag \\
\bar{v} & := \E v = \int_{v_{min}}^{v_{max}} v p(v) dv, \notag \\
\kappa & :=  \frac{1}{\delta \sqrt{\bar{v}}} \lim_{n \to \infty} \sqrt{\beta_n} \notag.
\end{align}
Note that if $p_n(\cdot)$ and $\beta_n$ do not vary with $n$ then $\kappa > 0$. If $\beta_n \to 0$ as $n \to \infty$, then $\kappa = 0$. Recall that for the random graph model described in Section \ref{sec:netmodel}, the off-diagonal entries of the adjacency matrix has mean and variance:
\begin{align*}
\E A_{ij} = &\rho w_i w_j,\\
\text{Var } A_{ij} = & \rho w_i w_j - (\rho w_i w_j)^2 \approx \rho w_i w_j.
\end{align*}
where $\rho = (1^T w)^{-1}$.  Define the following $n \times n$ matrix $C$:
\begin{align}
C
& := \frac{1}{\sqrt{n \rho}} W^{-1/2} \left( A - \rho w w^T \right) W^{-1/2} \notag\\
& = \sqrt{\frac{\beta_n}{n \mu}} V^{-1/2} \left( A - \frac{\mu v v^T}{\beta_n} \right) V^{-1/2} \label{eq:defC}.
\end{align}
It can be verified that $C$ is a standard Wigner matrix (\citet{taoBook}) where each off-diagonal entry has mean zero and variance $\frac{1}{n}$.  Now define the following $n \times n$ matrices:
\begin{align}
Y^{(1)}_n & := \frac{1}{n} \left(  V^{-1} - \sqrt{\frac{\beta_n}{\delta^2 \bar{v}}} C \right)^{-1},\\
Y^{(2)}_n & := \frac{1}{n} \left[ I - \sqrt{\frac{\beta_n}{\delta^2 \bar{v}}} V^{1/2} C V^{1/2}  \right]^{-1}, \\
Y^{(3)}_n & := \frac{1}{n} \left[ V^{1/2} \left( V^{-1} - \sqrt{\frac{\beta_n}{\delta^2 \bar{v}}} C \right)^{-1} V^{1/2}\right].
\end{align}
Note that the definitions of the matrices $Y_n^{(1)}, Y_n^{(2)}$  and $Y_n^{(3)}$ involve inverses of certain matrices. We argue later (in the proof of Lemma \ref{lemma:allTerms}) that the inverses exist since $M$ is stable almost surely. Using these expressions, we present a technical assumption required for our proof.
\begin{assumption}
\label{ass:expTrace}
For $k = 1, 2, 3$, suppose the following holds.
\begin{align}
	\lim_{n\to\infty}  \left[ 1^T \left( Y^{(k)}_n \right) 1  - \E \tr Y^{(k)}_n \right] = 0 \quad \as
\end{align}
\end{assumption}
\noindent Assumption \ref{ass:expTrace} essentially means that the sum of the off-diagonal entries of the matrices $Y_n^{(1)}, Y_n^{(2)}$  and $Y_n^{(3)}$ vanishes in the limit $n \to \infty$. We do not expect this assumption to be restrictive. All our simulations support this provided the system matrix $M$ is stable and we conjecture that the stability of $M$ is sufficient to guarantee the above. With this assumption and the notation presented above, we can now state the main result of this paper that calculates the disease cost $C_D(n)$:
\begin{theorem}
\label{thm:general} For $\left( G_{n, p_n(\cdot)}, \delta, \beta_n, \alpha, c_d \right)$, suppose $p_n(\cdot)$ has finite variance, the system matrices are almost surely stable for all $n$ and Assumption \ref{ass:expTrace} holds. Then
\begin{equation*}  \lim_{n \rightarrow \infty} C_D(n) =
	\begin{cases}
		\frac{\alpha c_d}{\delta} \left(  1 - \frac{\bar{v}^2}{\E v^2 - \delta \bar{v} } \right) \quad\as & \text{if } \kappa = 0,\\
		\frac{\alpha c_d}{\delta} \left(  1 + \kappa^2 F^2  - \frac{\kappa^2 F^2}{1 - \bar{v}/F - \delta \kappa^2 \bar{v}}  \right) \quad\as  &  \text{if } \kappa \neq 0.
	\end{cases}
\end{equation*}

\begin{equation*}
 \text{where } 	F = \int_{v_{min}}^{v_{max}} \frac{ p(v)}{ v^{-1} - \kappa^2 F } dv. \end{equation*}
\end{theorem}

Before we present the proof, we briefly remark on the assumptions required for the result to hold. The system matrix  $M = \delta I - \beta_n A$ is assumed to be almost surely stable. Essentially, this means that the disease dies out with high probability as the epidemic process proceeds on the random network. If this assumption does not hold, the cost $C_D (n)$ is infinite. We also assume the distribution $p_n(\cdot)$ has finite variance. To elucidate this assumption, suppose $\beta_n = \beta$ for all $n$ and the degree distribution $p_n(\cdot)$ is scale invariant. Thus $\kappa > 0$. In this regime, most degree distributions that do not have finite variance are heavy-tailed. It is well-known, e.g., \citet{Pastor01}, that over most networks with heavy-tailed degree distributions, there does not exist an infection threshold in the large graph limit, i.e., there is no positive ratio of $\delta/ \beta$ for which the infection dies out in these networks. Thus, since we require stability, we need not consider such networks. However, when the infection and network parameters scale with $n$, the connection between stability and finite variance is more involved; hence, we require both finite variance and stability.

For the remainder of this section, we focus on the proof of Theorem \ref{thm:general}.  We include a general overview of the proof in the following; most technical calculations are deferred to the appendix.

\textbf{Proof of Theorem \ref{thm:general}} \
The disease cost in \eqref{eq:diseaseCost} can be written as
\begin{align}
\lim_{n\to\infty} C_D(n) & = \lim_{n\to\infty} \frac{\alpha}{c_d} \left[ 1^T (I-M)^{-1} 1 \right] \notag\\
& =  \lim_{n\to\infty} \frac{\alpha c_d}{n} \left[ 1^T \left( \delta I - \beta {A} \right)^{-1} 1 \right]  \notag\\
& = \lim_{n\to\infty} \frac{\alpha c_d}{n} \left[ 1^T \left( \delta I - \beta_n \sqrt{n \rho} W^{1/2} C W^{1/2} - \beta \rho w w^T \right)^{-1} 1 \right] \notag\\
& = \lim_{n\to\infty} \frac{\alpha c_d}{n} \left[ 1^T \left( \underbrace{\delta I - \sqrt{n \beta_n \mu} V^{1/2} C V^{1/2}}_{:=X} - \mu v v^T \right)^{-1} 1 \right] \label{eq:defX}\\
& =  \lim_{n\to\infty} \frac{\alpha c_d}{n} \left[ 1^T \left( X - \mu v v^T \right)^{-1} 1 \right] \notag.
\end{align}
Applying the Matrix Inversion Lemma (\citet{horn2005matrix}), we get
\begin{align} \lim_{n\to\infty} C_D(n) & =  \lim_{n\to\infty} \frac{\alpha c_d}{n} \left[ 1^T \left( X^{-1} - \frac{X^{-1} v v^T X^{-1}}{-\frac{1}{\mu} + v^T X^{-1} v} \right) 1 \right]  \notag \\
& = \alpha c_d \lim_{n\to\infty} \left[ \left( \frac{1}{n} 1^T X^{-1} 1 \right) - \frac{ \left( \frac{1}{n} 1^T X^{-1} v \right)^2}{-\frac{1}{n \mu} + \left( \frac{1}{n} v^T X^{-1} v \right)} \right]. \label{eq:calc}
\end{align}

From the Strong Law of Large Numbers, we have $\lim_{n \to \infty} {n \mu} = 1/\bar{v}$ almost surely. To proceed, we show that each of the terms in \eqref{eq:calc}, i.e., $\frac{1}{n} ( 1^T X^{-1} 1)$, $\frac{1}{n} (1^T X^{-1} v )$, and $\frac{1}{n} (v^T X^{-1} v)$, almost surely self-average under certain technical conditions (Assumption \ref{ass:expTrace}) and can be computed easily  using $p_n(\cdot)$, $\kappa$ and $\delta$.   The result of this computation is summarized as follows:
\begin{lemma}
\label{lemma:allTerms}
If Assumption \ref{ass:expTrace} holds, then
\begin{align}
	& \text{(a)} \lim_{n \to \infty} \frac{1}{n} 1^T X^{-1} v  = \frac{F}{\delta}   \quad\text{a.s.} \label{eq:1xw}\\
	& \text{(b)} \lim_{n \to \infty} \frac{1}{n} 1^T X^{-1} 1  =  \frac{1 + \kappa^2 F^2}{\delta}   \quad\text{a.s.} \label{eq:1x1}\\
	& \text{(c)} \lim_{n \to \infty} \frac{1}{n} v^T X^{-1} v = \begin{cases}
	\E v^2 \quad\text{a.s.} & \text{if } \kappa = 0,\\
	\frac{1}{\delta \kappa^2} \left( 1- \frac{\bar{v}}{F} \right) \quad\text{a.s.} & \text{if } \kappa \neq 0.
	\end{cases}\label{eq:wxw}
\end{align}
\noindent where $F$ is the solution of the following fixed point equation:
\begin{align}
\label{eq:defF}
	F = \int_{v_{min}}^{v_{max}} \frac{ p(v)}{ v^{-1} - \kappa^2 F } dv.
\end{align}
\end{lemma}
\noindent We defer the proof of Lemma \ref{lemma:allTerms} to Appendix \ref{sec:appProofs}.  Using this lemma and the fact that $\lim_{n \rightarrow \infty} n\mu = 1/\bar{v} \as$ in \eqref{eq:calc}, we have, for $\kappa \neq 0$:
\begin{align*}
\lim_{n\to\infty}  C_D(n)
& = \alpha c_d \lim_{n\to\infty} \left[ \left( \frac{1}{n} 1^T X^{-1} 1 \right) - \frac{ \left( \frac{1}{n} 1^T X^{-1} v \right)^2}{-\frac{1}{n \mu} + \left( \frac{1}{n} v^T X^{-1} v \right)} \right]  \\
& = \alpha c_d \left( \frac{1 + \kappa^2 F^2}{\delta} - \frac{ \frac{F^2}{\delta^2}}{   - \bar{v} + \frac{1}{\delta \kappa^2} \left( 1- \frac{\bar{v}}{F} \right)} \right)\as \\
& = \frac{\alpha c_d}{\delta} \left(  1 + \kappa^2 F^2  - \frac{\kappa^2 F^2}{1 - \bar{v}/F - \delta \kappa^2 \bar{v}}  \right) \as \end{align*}
Similarly, the case for $\kappa = 0$ follows by substituting the relevant expressions from Lemma \ref{lemma:allTerms} in \eqref{eq:calc}, which completes the proof. $\square$

\subsection{Bounds for a fixed network} \label{subsec:boundFixedGraph}
Previously we considered the asymptotic regime for the epidemic process $\left( G_{n, p_n(\cdot)}, \delta, \beta_n, \alpha, c_d \right)$ and computed the cost exactly for a specific scaling of the parameters. In this section, we fix the graph and compute a bound on the cost of the epidemic process $\left( G, \delta, \beta, \alpha, c_d \right)$. Since the bound applies to any specific instance of the graph, it also applies to a family of graphs generated according to the random graph model in Section \ref{sec:netmodel}.


\begin{theorem} \label{thm:bound} For $\left( G, \delta, \beta, \alpha, c_d \right)$, with a stable system matrix $M = (1-\delta) I - \beta A$, the cost of disease per node satisfies
\begin{equation*} C_D (n) \leq \frac{\alpha c_d}{1 - \lambda_{max} ( M )} \end{equation*}
where $\lambda_{max}(.)$ denotes the maximum eigenvalue of the corresponding matrix.
\end{theorem}
\proof{Proof.}
For any matrix $H$, it is said to be positive definite (or positive semidefinite) if the eigenvalues of $H$ are strictly positive (or non-negative, respectively). Further, it is denoted as $H \succ 0$ (or $H\succeq 0$, respectively). Since $M$ is stable, we have $(I-M) \succ 0$ and $\lambda_{max}(M) < 1$. Then, it follows that
\begin{align*}
(I-M)^{-1}  \ & \preceq \ \frac{1}{1- \lambda_{max}(M)} I\\
\implies 1^T (I-M)^{-1} 1 \ & \leq \ 1^T \left( \frac{ I }{1-\lambda_{max}(M)} \right) 1
\end{align*}
The rest follows from the definition of $C_D(n)$ in \eqref{eq:diseaseCost}.
\Halmos
\endproof

We note that the necessary and sufficient condition required for the disease to die out and the social cost to converge are the same, i.e., $\lambda_{max}(M) < 1$. Also, note that the bound only depends on $\lambda_{max}(M)$ from the disease propagation model, which is popularly known as the disease threshold.  It is interesting that the same parameter of the disease plays the central role in both the tapering off of the disease and its total cost.


\subsection*{Example: The Erd\"{o}s-R\'{e}nyi network} \label{sec:ER}
To illustrate the tightness of the bound, we consider a specific type of network -- the Erd\"{o}s-R\'{e}nyi random graph. As described in Example \ref{ex:ER}, in this type of network, an edge exists between each pair of nodes with uniform probability $p$. To generate this graph using our network model, the degree distribution $p_n(\cdot)$ is a delta function at $np$. It is well-known that such graphs are connected with high probability if $p > \log n / n$ in the large graph regime, e.g., \citet{chungBook}. We study the cost of disease in this type of network when the infection rate $\beta_n$ scales such that $\beta_n n p$ is a constant. Since we are interested in the regime $np \to \infty$, the infection rate satisfies $\beta_n \to 0$ and $\kappa = 0$ in this case. The scale invariant distribution $p_n(\cdot)$ has a bounded support and satisfies:
\begin{align*}
p(v) & = \boldsymbol{\delta}(v - \beta_n n p ),\\
\E v & = \bar{v} = \beta_n n p, \\
\E v^2 & = (\beta_n n p)^2.
\end{align*}
\noindent We refer to the constant $\beta_n n p$ as $\bar{v}$. Using this notation, Theorem \ref{thm:general} yields
\begin{align}
\lim_{n \to \infty} C_D (n)
& = \frac{\alpha c_d}{\delta} \left(  1 - \frac{\bar{v}^2}{\E v^2 - \delta \bar{v} } \right) \quad\text{a.s.}  \notag \\
& = \frac{\alpha c_d}{\delta} \left(  1 - \frac{\bar{v}^2}{\bar{v}^2 - \delta \bar{v} } \right) \quad\text{a.s.} \notag \\
& = \frac{\alpha c_d}{\delta - \bar{v}} \quad\text{a.s.} \label{eq:erExactCost}
\end{align}
Note that this is essentially the same result as in \citet{gameNets11}, but applied to the case where the network and infection parameters scale with $n$.

Now we illustrate the bound in Theorem \ref{thm:bound} using this class of networks. Consider a randomly sampled Erd\"{o}s-R\'{e}nyi graph with a reasonably large population size $n$ and edge-forming probability $p$. To ensure connectedness with high probability, $p$ is chosen to be greater than $\log n/ n$. The infection parameters $\delta$ and $\beta$ are selected such that $\delta > \beta n p$. As a result, the maximum eigenvalue of the adjacency matrix of this graph ($\lambda_{max}[A]$) is $np$ with high probability (denoted as w.h.p.), e.g., in \citet{chungBook}. Thus we have
\begin{align*}
\lambda_{max}(M) = 1 - \delta + \beta \quad\text{w.h.p.}
\end{align*}
The choice of $\delta > \beta n p$ ensures that $M$ is stable w.h.p. Using Theorem \ref{thm:bound}, we then have
\begin{align}
C_D (n) \leq \frac{\alpha c_d}{\delta - \beta n p} \quad\text{w.h.p.} \label{eq:erBoundCost}
\end{align}
It is interesting that for an Erd\"{o}s-R\'{e}nyi network of large population size the exact cost in \eqref{eq:erExactCost} as calculated from Theorem \ref{thm:general} coincides exactly with the bound in \eqref{eq:erBoundCost} as calculated from Theorem \ref{thm:bound}. This can be explained as follows. For an Erd\"{o}s-R\'{e}nyi network, the adjacency matrix has a large maximum eigenvalue $(np)$ as compared to the rest of the spectrum that is concentrated in the interval $[-2 \sqrt{np}, 2 \sqrt{np}]$ from Wigner's Semicircle Law, as in \citet{taoBook}. The system matrix $M$ has an eigen-spectrum that is a translation and stretch of the eigen-spectrum of $A$. Thus it also has the property that $\lambda_{max}(M) \gg \lambda(M)$, where $\lambda(M)$ is a randomly sampled eigenvalue of $M$. The calculations from Theorem \ref{thm:general} and \ref{thm:bound} coincide since the largest eigenvalue dominates over the other eigenvalues.  We illustrate the closeness of the disease cost as predicted by the two theorems for an Erd\"{o}s-R\'{e}nyi network using simulations in the next section. 

%% file: sim_disc.tex
\section{Numerical simulations}
\label{sec:sim_disc}

In this section we use simulations to illustrate the results in the previous sections. In Section \ref{subsec:evalAss}, we explore how accurate the linearized model of epidemic process in \eqref{eq:linModel} is when contrasted with the actual disease propagation. According to our simulations, when the infection rate $\beta$ is small enough and the degree distribution $p_n (.)$ is not too heavy in its tail, the cost computed via the linear model of infection spread is a good estimate of the actual disease cost.  In the second part of this section, we illustrate Theorems \ref{thm:general} and \ref{thm:bound} in Section \ref{subsec:evalThm}, simulating the disease on both random and real-world networks and comparing the results to the analytical results from our theorems. Note that the analytical results of the theorems have been derived under certain technical assumptions; the simulations are performed to compare the robustness of these results even in settings, where some of the assumptions do not apply.

\subsection{Evaluation of assumptions}
\label{subsec:evalAss}
\begin{figure}[htb] \centering
 \subfloat[Percent Error]{\label{fig:errorER}\includegraphics[width=0.5\textwidth]{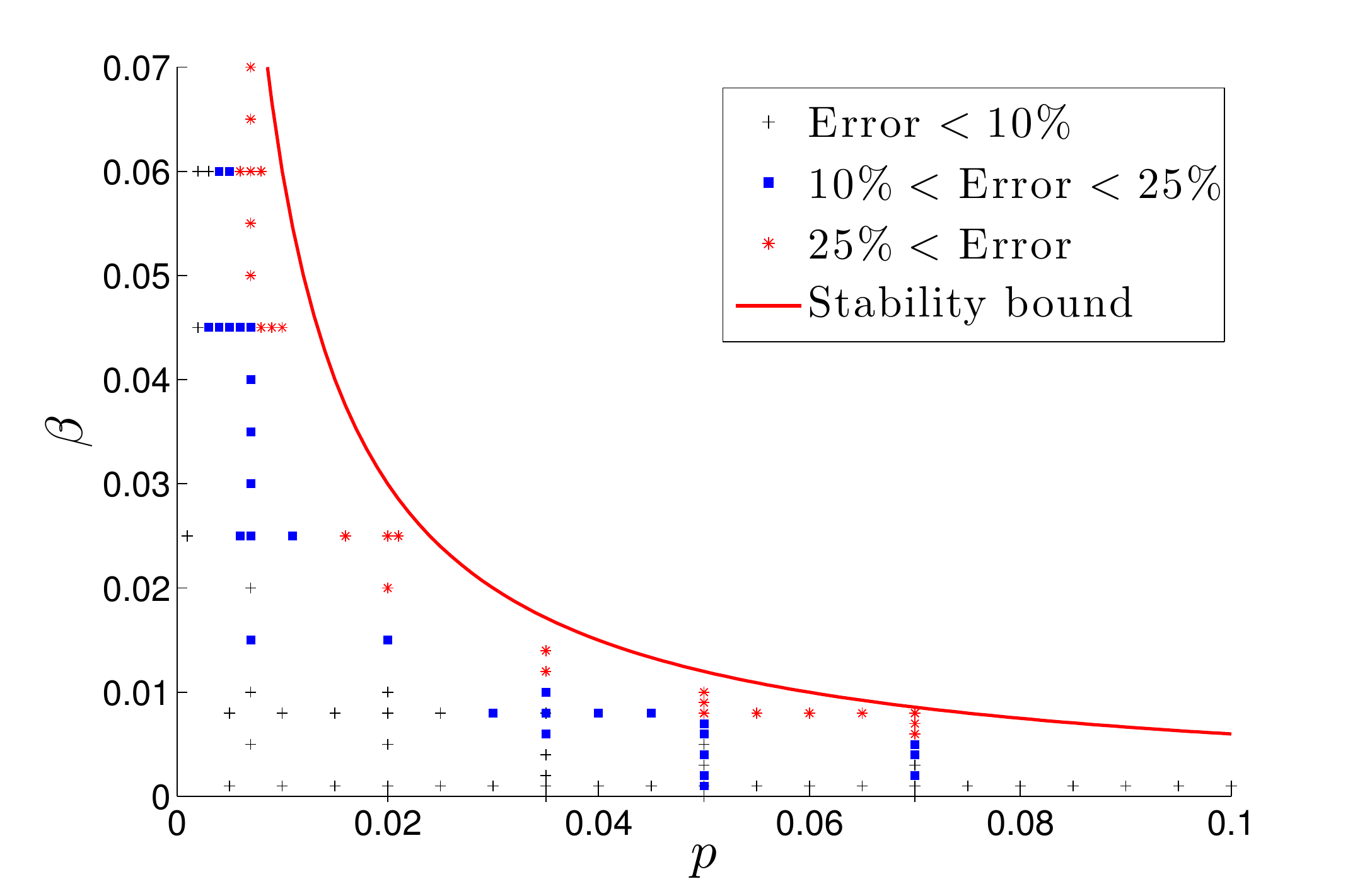}}
 \subfloat[Estimated bounds for percent error]{\label{fig:linesER}\includegraphics[width=0.5\textwidth]{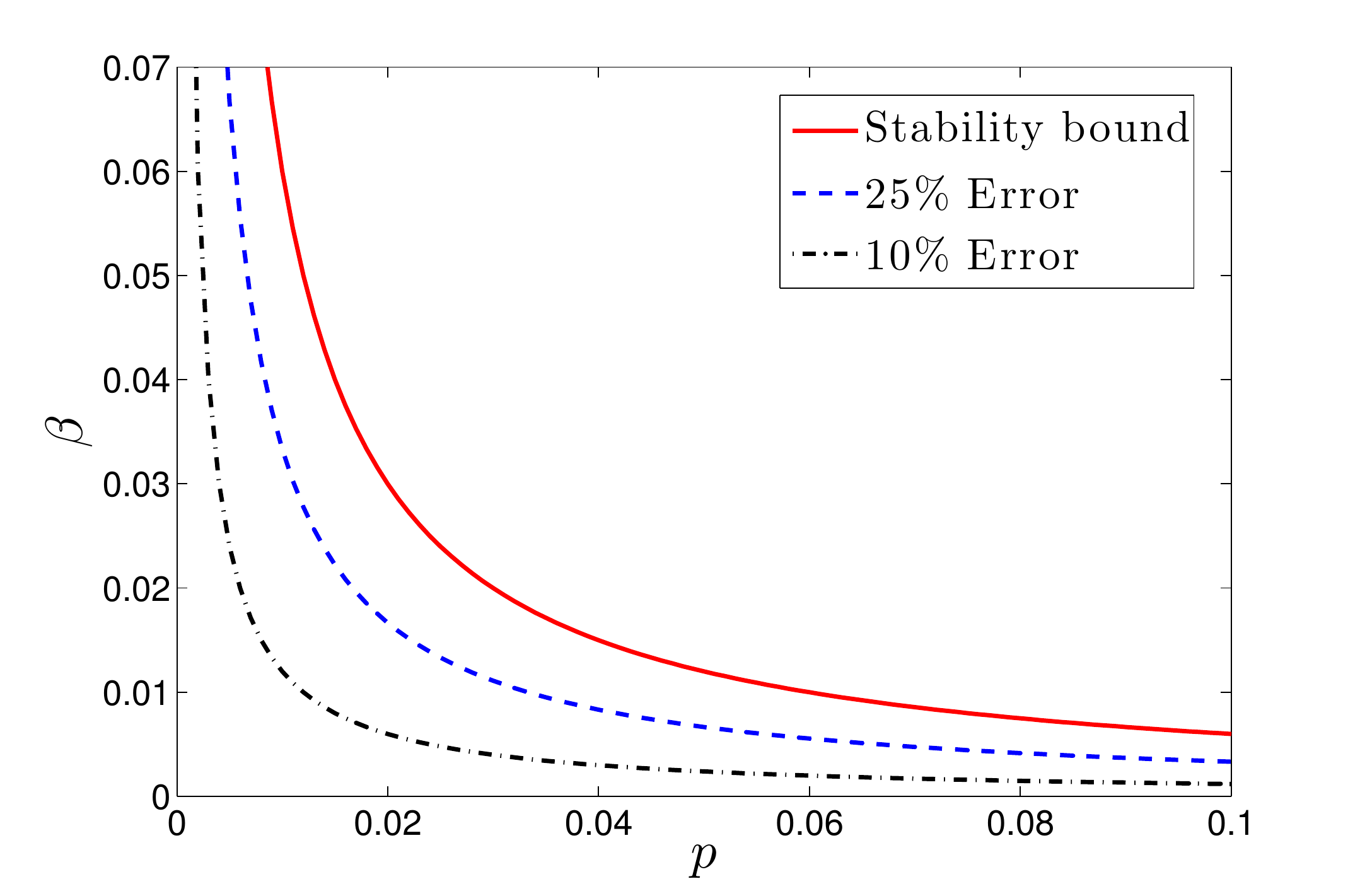}}
  \caption{Percent error between simulated cost and linearized model \eqref{eq:diseaseCost} for ER network}
  \label{fig:ERerror}
\end{figure}

To analyze the parameter regimes for which the linear disease propagation model is accurate, first consider an Erd\"{o}s-R\'{e}nyi network with $n = 1000$. Assume the recovery rate is $\delta = 0.6$ and the initial fraction of infected nodes is $\alpha = 0.2$. We evaluate two quantities: (i) $\alpha c_d [1^T (I - M)^{-1} 1]$ and (ii) actual cost by summing up the number of infected nodes at each time until the disease dies out, normalizing to obtain the average per-node cost. We calculate the relative error between the linear model and the actual cost, averaged over 100 runs. To determine the parameter regimes for which our model is accurate, we simulate with various values of the infection rate $\beta$ and edge probability $p$. Our results are presented in Figure \ref{fig:ERerror}. Note that in the case of an  Erd\"{o}s-R\'{e}nyi random network,  stability of the system matrix requires that $\delta > \beta n p$; this bound is shown in Figure \ref{fig:ERerror}. Outside of this bound, when both $\beta$ and $p$ are large, the disease does not die out. Hence we determine regions that guarantee error percentages within certain ranges. Note that much of the area within the stability region has relative error less that $10\%$, indicating that the approximation is good in regimes where the disease dies out.

\begin{figure}[htb] \centering
 \subfloat[Percent Error]{\label{fig:errorPL}\includegraphics[width=0.5\textwidth]{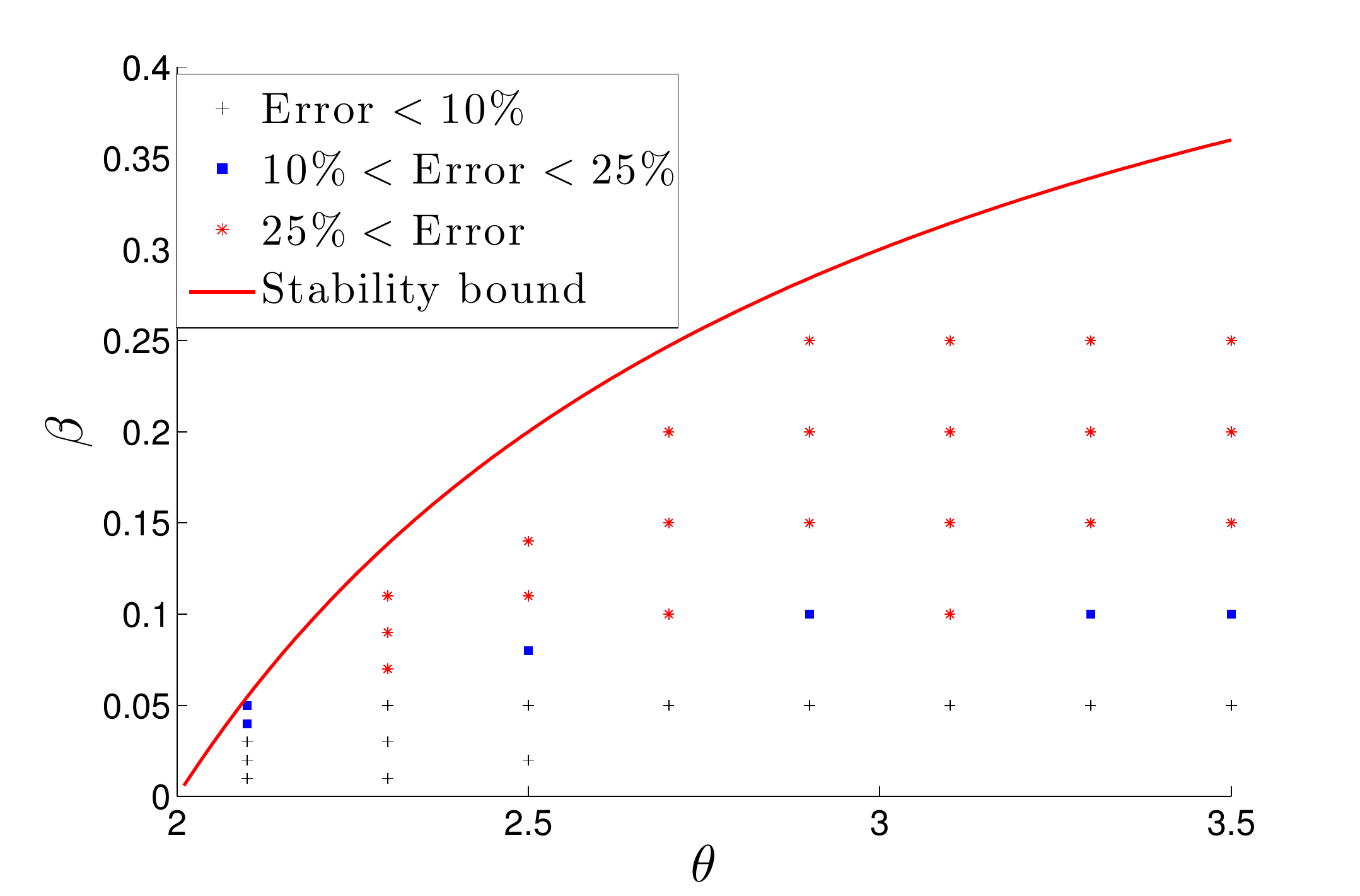}}
 \subfloat[Estimated bounds for percent error]{\label{fig:linesPL}\includegraphics[width=0.5\textwidth]{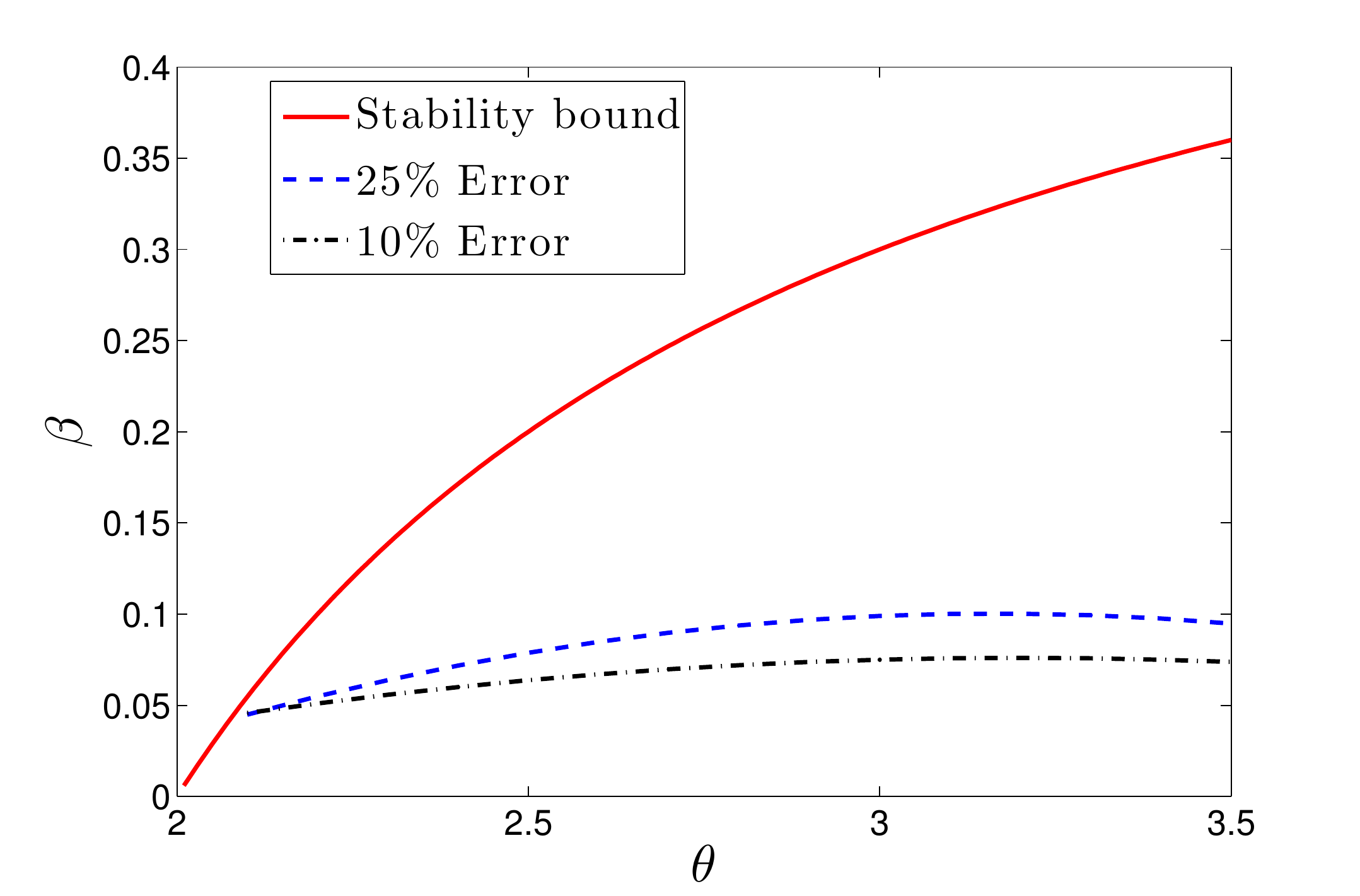}}
  \caption{Percent error between simulated cost and linearized model \eqref{eq:diseaseCost} for Pareto network}
  \label{fig:PLerror}
\end{figure}

Now consider a 1000-node network with a Pareto degree distribution, as described in Example \ref{ex:PL}. Note that this is a heavy-tailed distribution and only has finite variance for $\theta > 2$. Again, we test various network and infection parameters (in this case, $\beta$ and $\theta$) to determine the relative error between our linearized model and the actual epidemic process.  The results are presented in Figure \ref{fig:PLerror}. As expected, a heavier tail (smaller $\theta$) results in a larger relative error. Similarly, a higher infection rate $\beta$ results in a larger error. However, there does exist a decent sized region within which the linear model is a good approximation.

The Erd\"{o}s-R\'{e}nyi random network and a network with a heavy tailed distribution represent the extreme cases, in terms of degree distribution. Where an Erd\"{o}s-R\'{e}nyi network has a degree distribution that is concentrated around the mean, the heavy-tailed Pareto distribution is characterized by large deviations in the degrees of each node. Our simulations show that under reasonable assumptions on the graph parameters and the infection rate, the linear model is quite accurate wherever the disease dies out (inside the stability region), thus validating our choice of using it to compute the costs in Theorems \ref{thm:general} and \ref{thm:bound}.

\subsection{Illustration of theorems}
\label{subsec:evalThm}
In this section we compare our different expressions for the disease cost, from \eqref{eq:diseaseCost}, Theorem \ref{thm:general} and the bound in Theorem \ref{thm:bound} for various types of random and real-world graphs.  We show through simulations that despite the approximations made in calculating the closed form solution in Theorem \ref{thm:general}, it is very close to the original expression of the disease cost from \eqref{eq:diseaseCost}, as well to the simulated spread of the the disease.  Further, for some types of graphs, the bound in Theorem \ref{thm:bound} is also rather tight.  Note that in all cases, as $n$ grows, our approximations become tighter.

\begin{figure}[htb] \centering
 \subfloat[$n=100$]{\label{fig:thmER100}\includegraphics[width=0.5\textwidth]{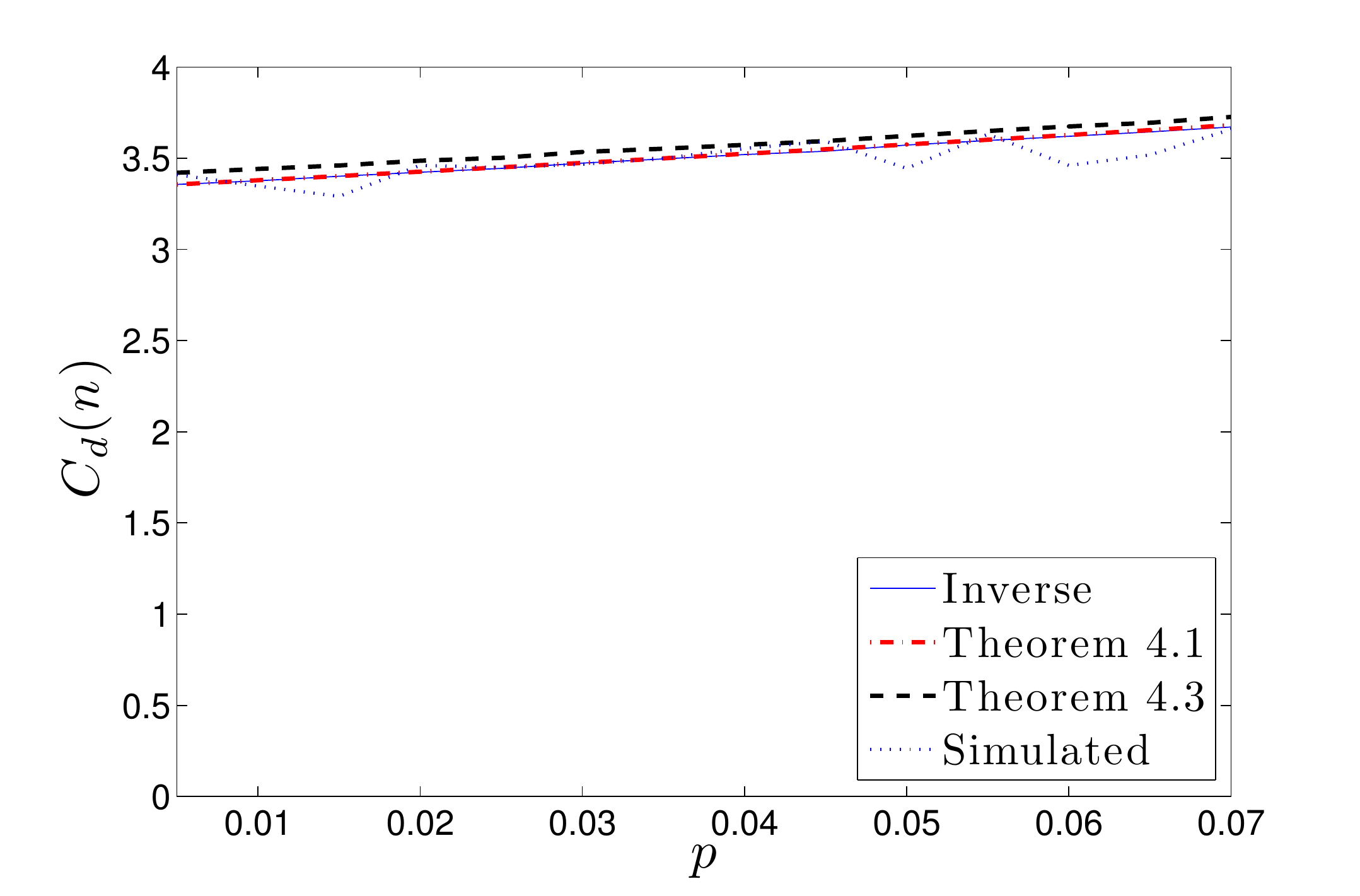}}
 \subfloat[$n=1000$]{\label{fig:thmER1000}\includegraphics[width=0.5\textwidth]{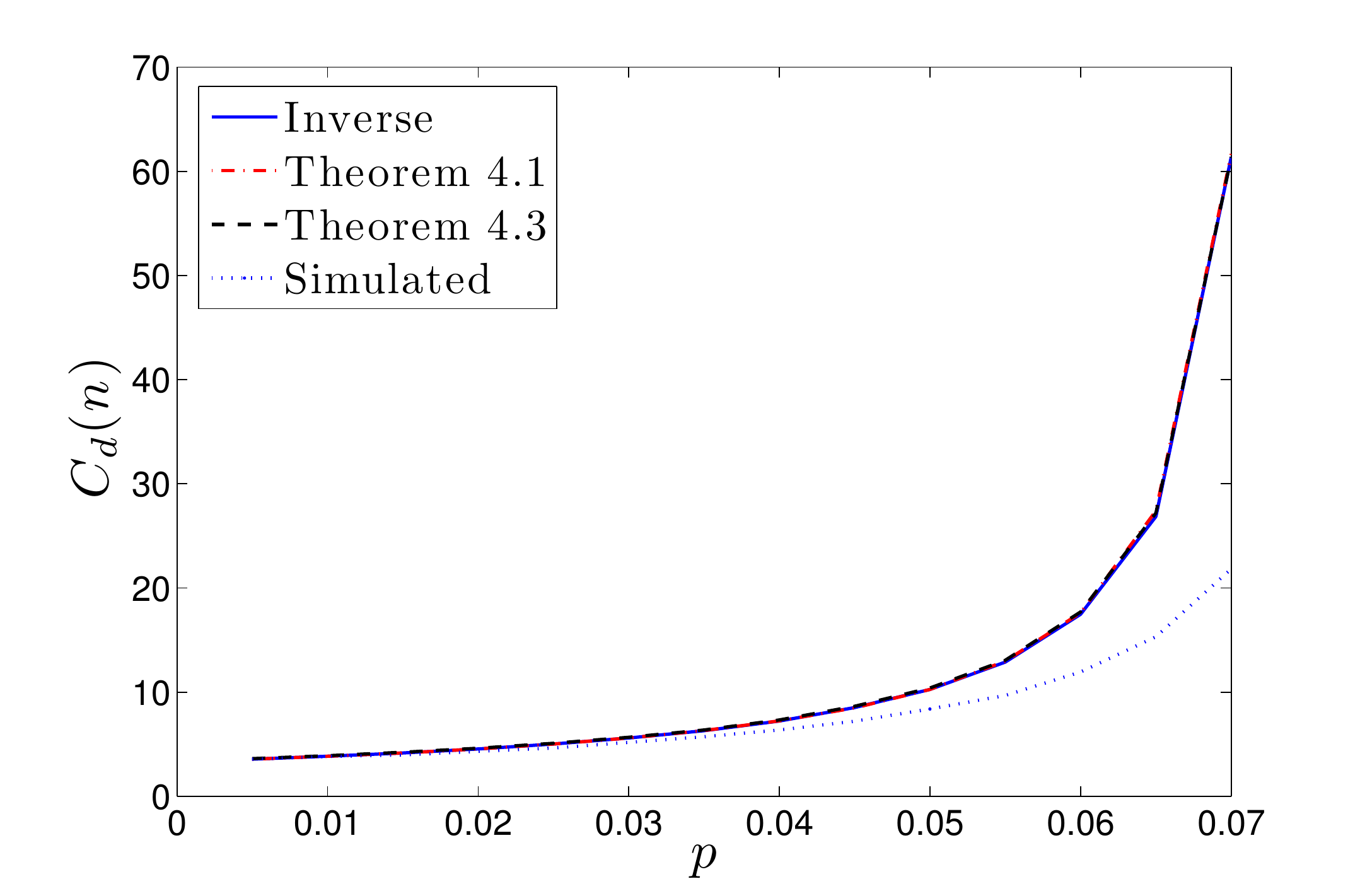}}
  \caption{Simulated and calculated disease cost on ER network}
  \label{fig:thmER}
\end{figure}

As expected, the Erd\"{o}s-R\'{e}nyi random network has the closest agreement between Theorems \ref{thm:general} and \ref{thm:bound} and the linear model, from \eqref{eq:diseaseCost}.  Similar to the accuracy regimes described above, as the expected degree grows, the gap between the linear model and the actual simulation increases, as shown in Figure \ref{fig:thmER}.

\begin{figure}[htb] \centering
 \subfloat[$n=100$]{\label{fig:thmExp100}\includegraphics[width=0.5\textwidth]{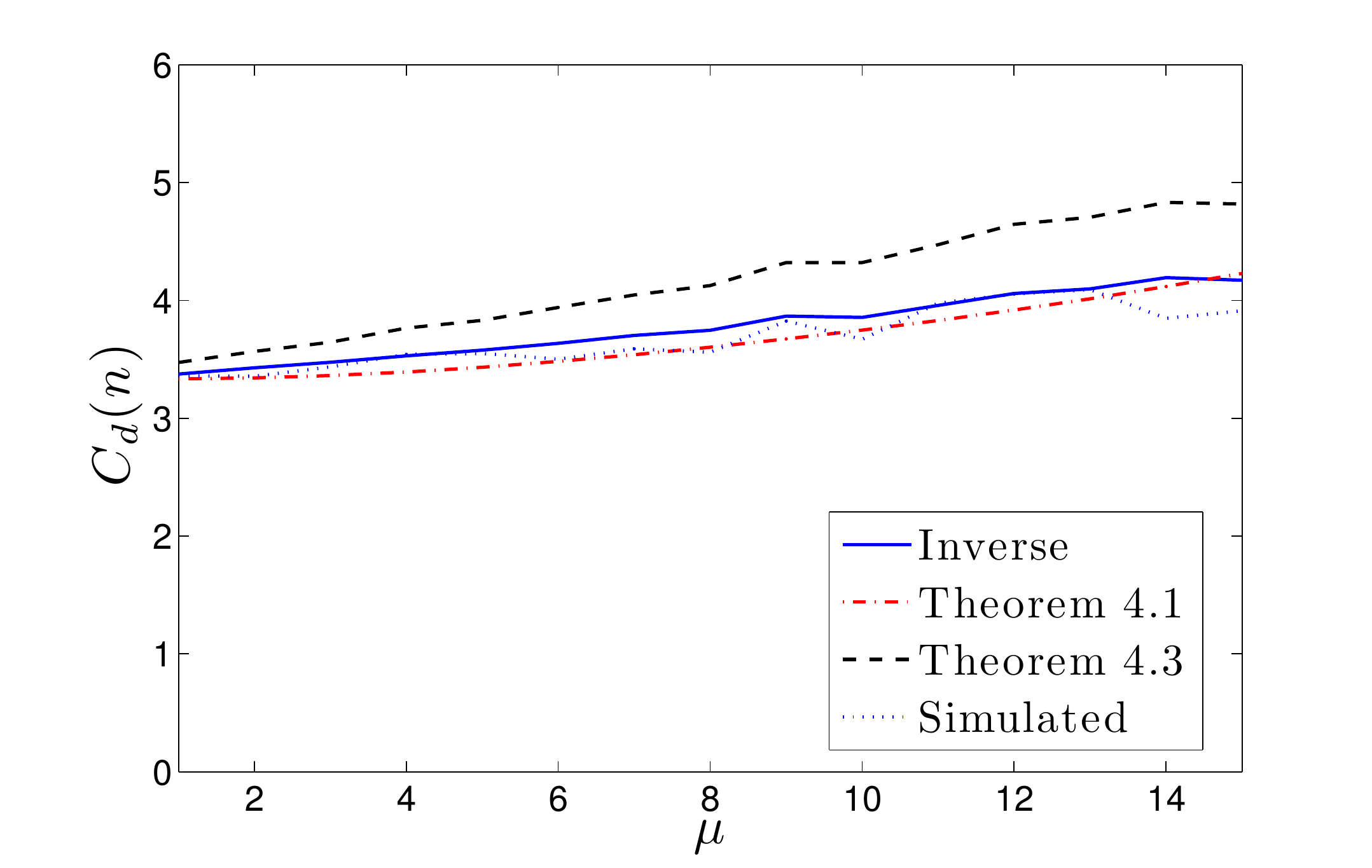}}
 \subfloat[$n=1000$]{\label{fig:thmExp1000}\includegraphics[width=0.5\textwidth]{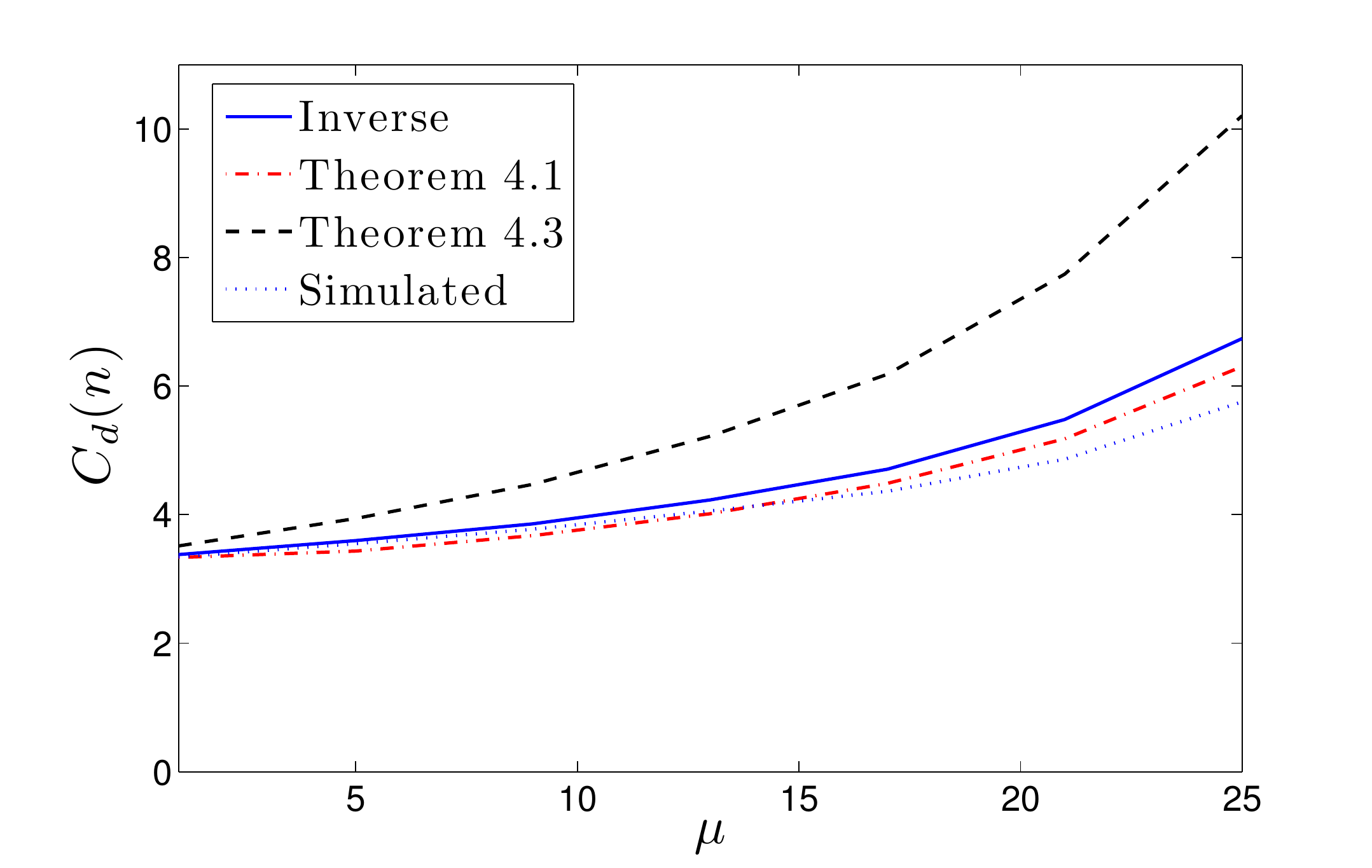}}
  \caption{Simulated and calculated disease cost on exponential network}
  \label{fig:thmExp}
\end{figure}

For a network with an exponential degree distribution, as shown in Figure \ref{fig:thmExp}, we see fairly close agreement between Theorem \ref{thm:general}, the linear model \eqref{eq:diseaseCost}, and the actual simulation.  However, the upper bound from Theorem \ref{thm:bound} is less tight than in the Erd\"{o}s-R\'{e}nyi case.

\begin{figure}[htb] \centering
 \subfloat[$n=100$]{\label{fig:thmPL100}\includegraphics[width=0.5\textwidth]{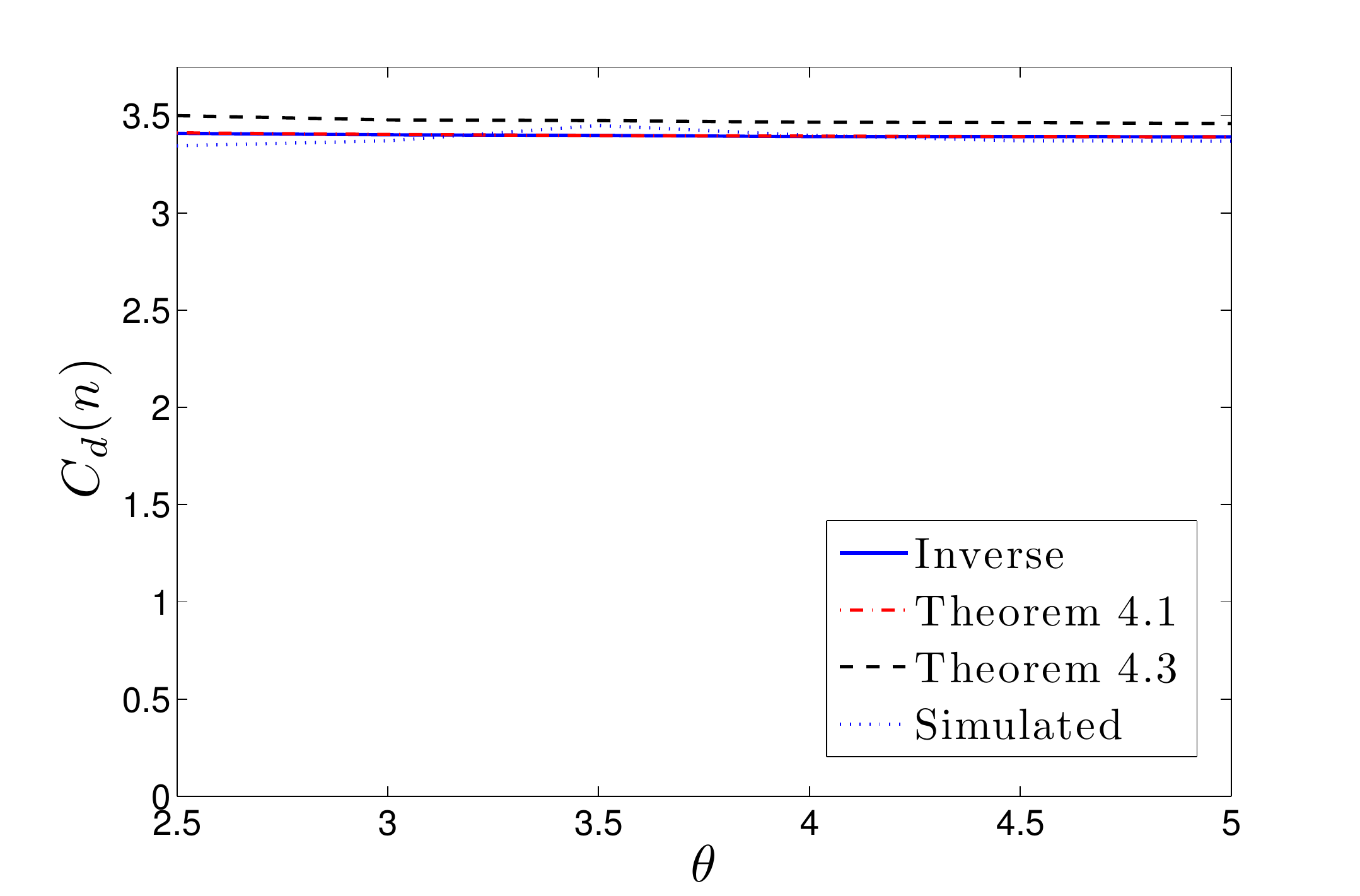}}
 \subfloat[$n=1000$]{\label{fig:thmPL1000}\includegraphics[width=0.5\textwidth]{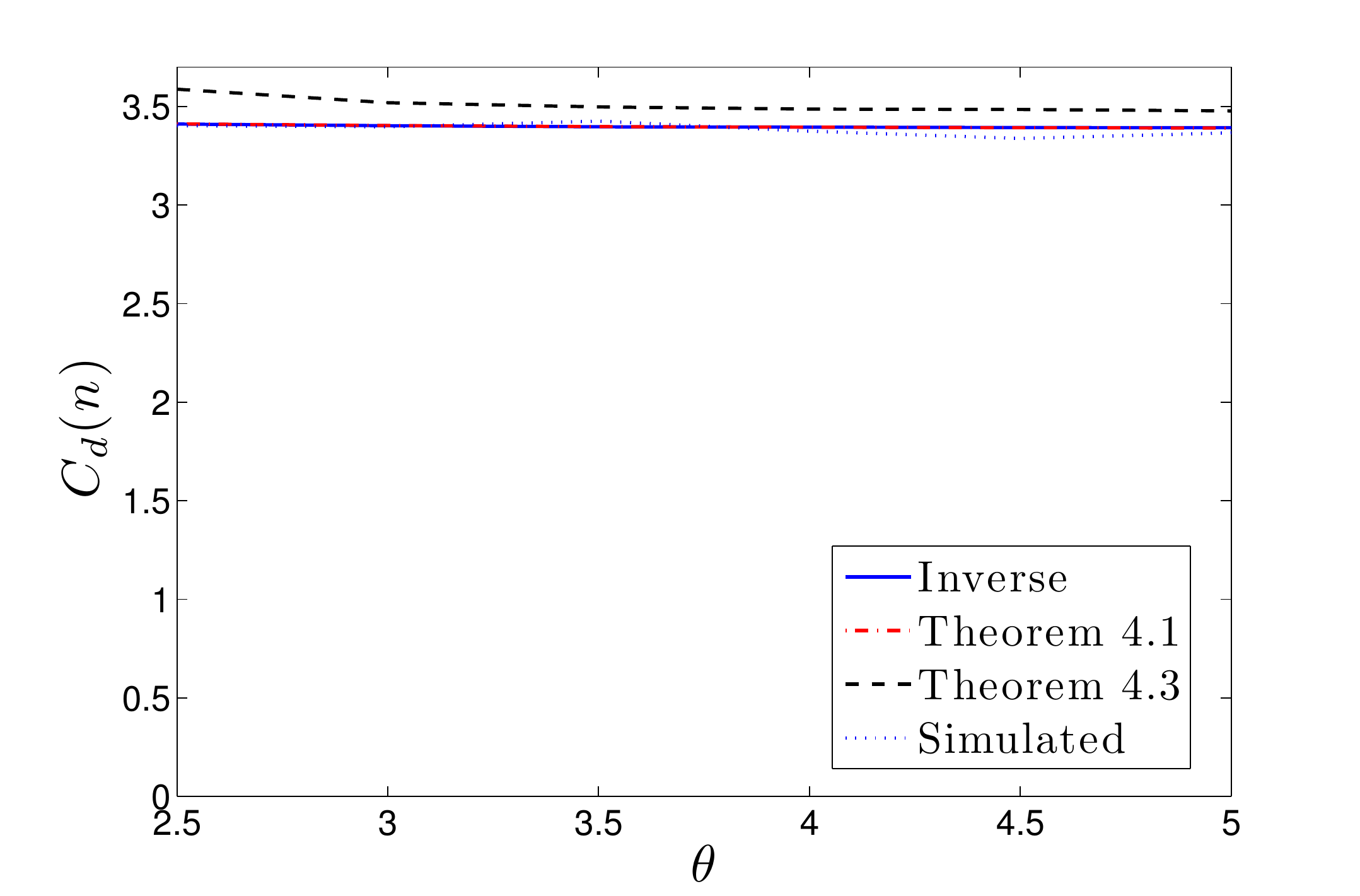}}
  \caption{Simulated and calculated disease cost on Pareto network}
  \label{fig:thmPL}
\end{figure}

The network with a Pareto degree distribution has similar results to one with an exponential degree distribution.  Again, we see close agreement between the linear model and Theorem \ref{thm:general}.  However, due to the increased variance in node degrees, the simulated cost of diseases varies more than in the networks with a more concentrated degree distribution.

\subsection{Disease cost case studies}
\begin{figure}[htb] \centering
 \subfloat[Network]{\label{fig:caltechNW}\includegraphics[width=0.5\textwidth]{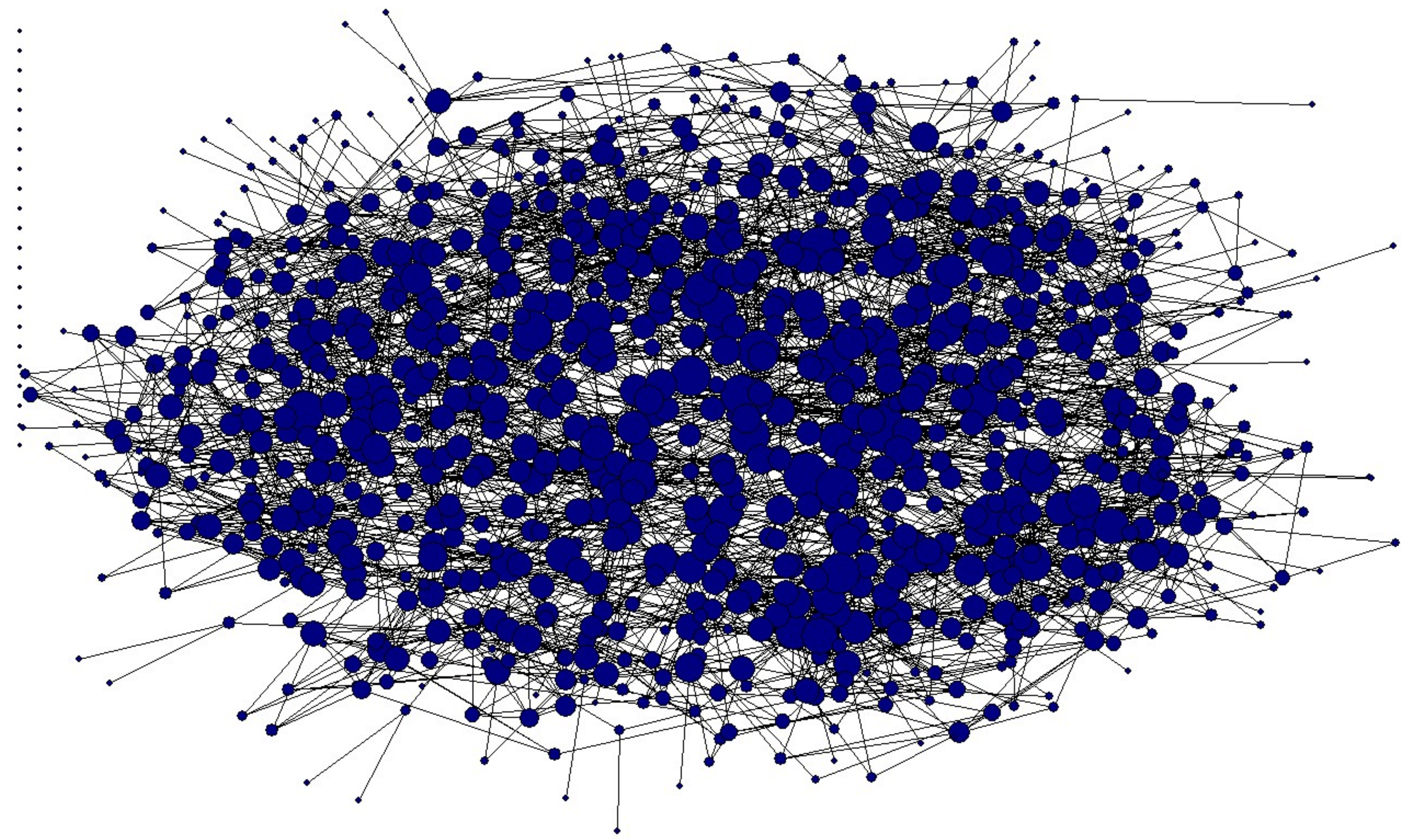}}
 \subfloat[CCDF of degree distribution]{\label{fig:caltechDeg}\includegraphics[width=0.5\textwidth]{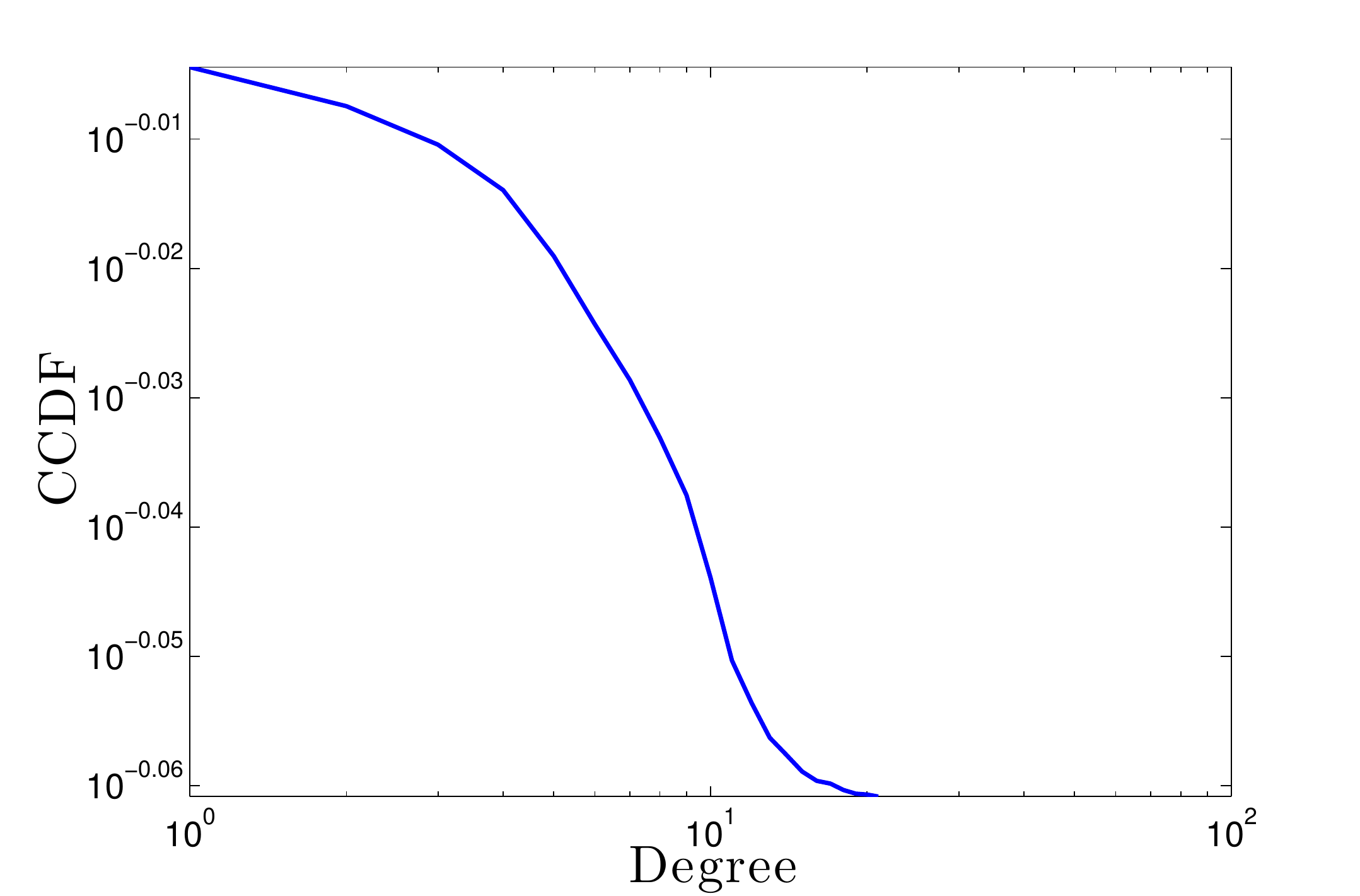}}
  \caption{Caltech social network}
  \label{fig:caltechSN}
\end{figure}

To illustrate our results on actual networks, we present two case studies here, evaluating Theorem \ref{thm:bound} and simulating the disease on each.  The first real-world network we examine is a social network of undergraduates at the California Institute of Technology (Caltech) in \citet{caltechProject}.  This data was gathered via a survey in 2010, asking participating students to list up to 10 of their friends.  Participation was about $72\%$ of the undergraduate student body, resulting in at least partial network information for about $95\%$ of the students.  We generate an undirected network for our simulations by making each directed edge undirected.  The final network has about $900$ nodes and $3500$ edges.  Both the network and its degree distribution (the CCDF in a $\log\log$ plot) are shown in Figure \ref{fig:caltechSN}.

\begin{figure}[h!] \centering
 \subfloat[Network]{\label{fig:wikiNet}\includegraphics[width=0.5\textwidth]{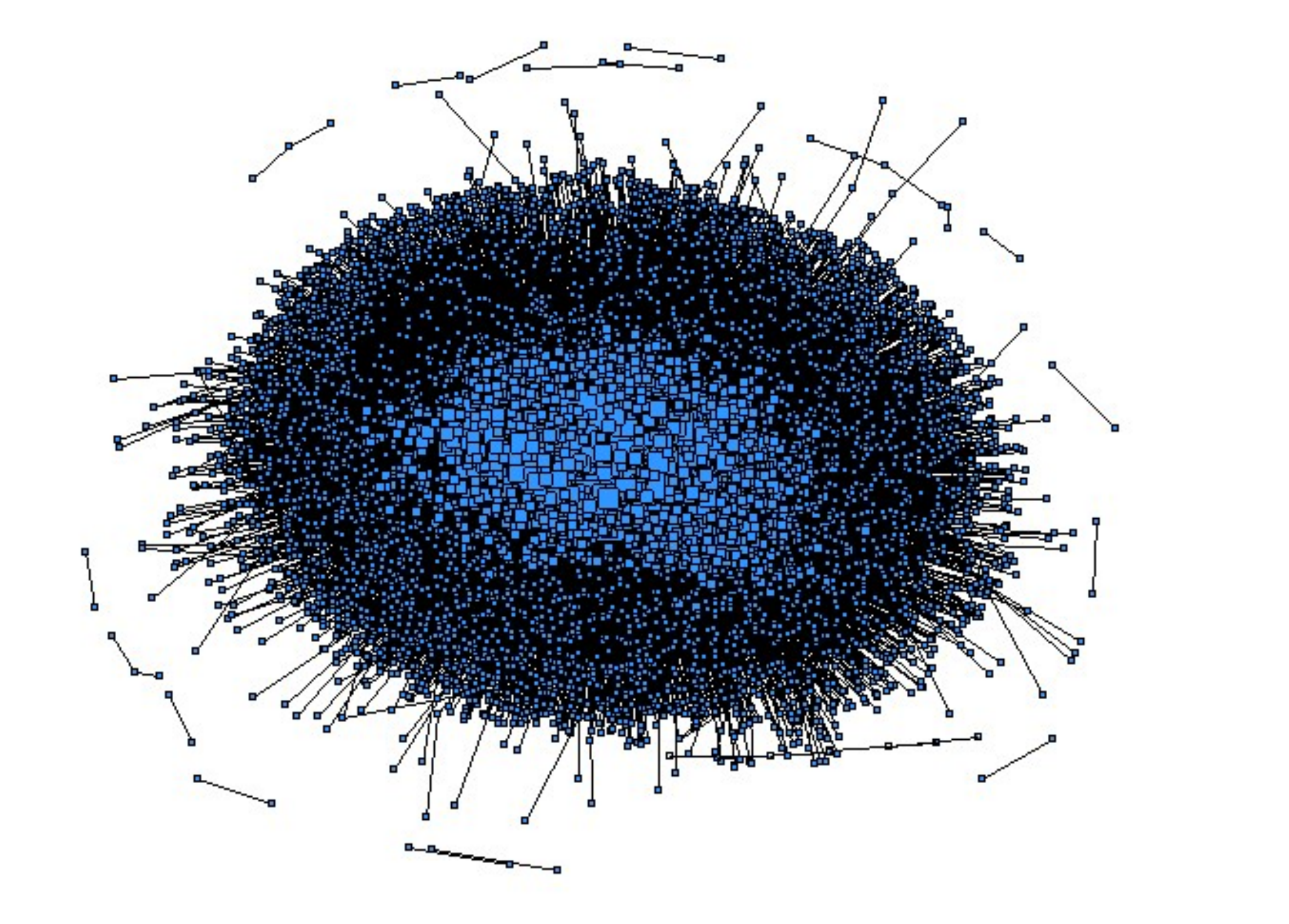}}
 \subfloat[CCDF of degree distribution]{\label{fig:wikiDeg}\includegraphics[width=0.5\textwidth]{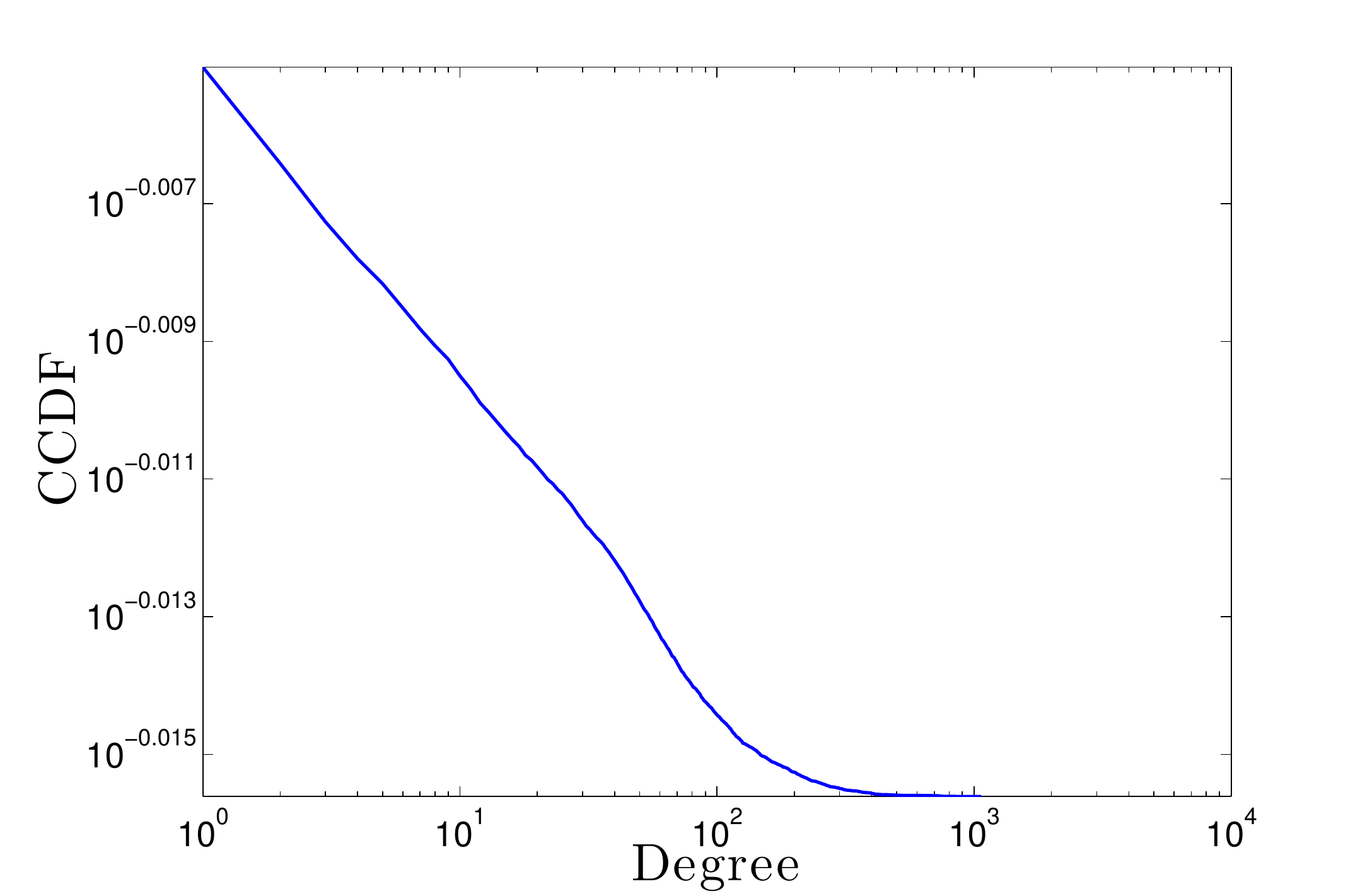}}
  \caption{Wikipedia voting network}
  \label{fig:wikiNW}
\end{figure}

The second network we look at is gathered from voting records in Wikipedia\footnote{Datasets available at: \url{http://snap.stanford.edu/data/wiki-Vote.html}.}. This network has about $7000$ nodes and $100000$ edges.  The network and its degree distribution (the CCDF in a $\log\log$ plot) are shown in Figure \ref{fig:wikiNW}.  Though this is not an actual social network, it is a good representation of a larger data set with similar characteristics to social networks.

\begin{figure}[h!] \centering
 \subfloat[Caltech]{\label{fig:caltechSim}\includegraphics[width=0.5\textwidth]{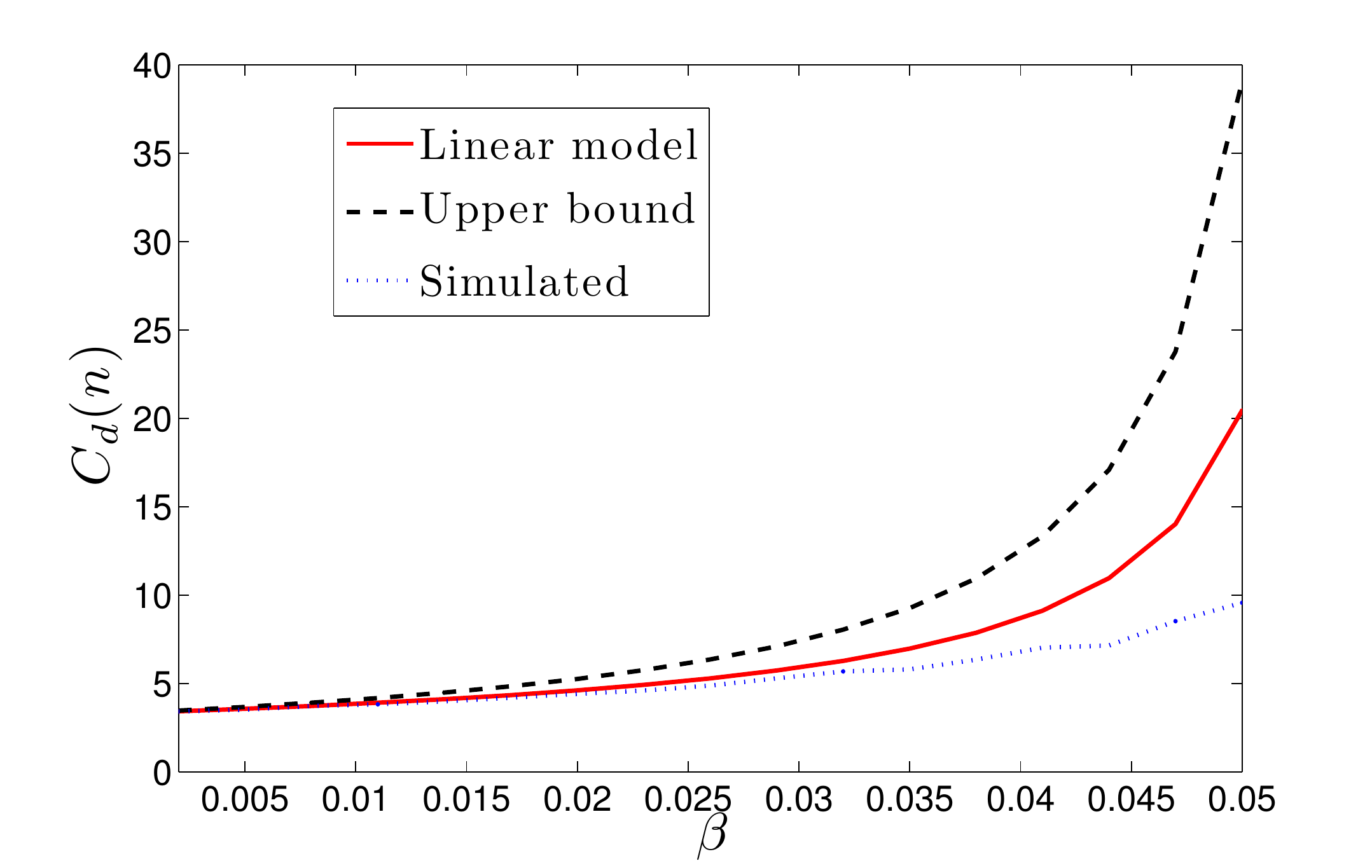}}
 \subfloat[Wikipedia]{\label{fig:wikiSim}\includegraphics[width=0.5\textwidth]{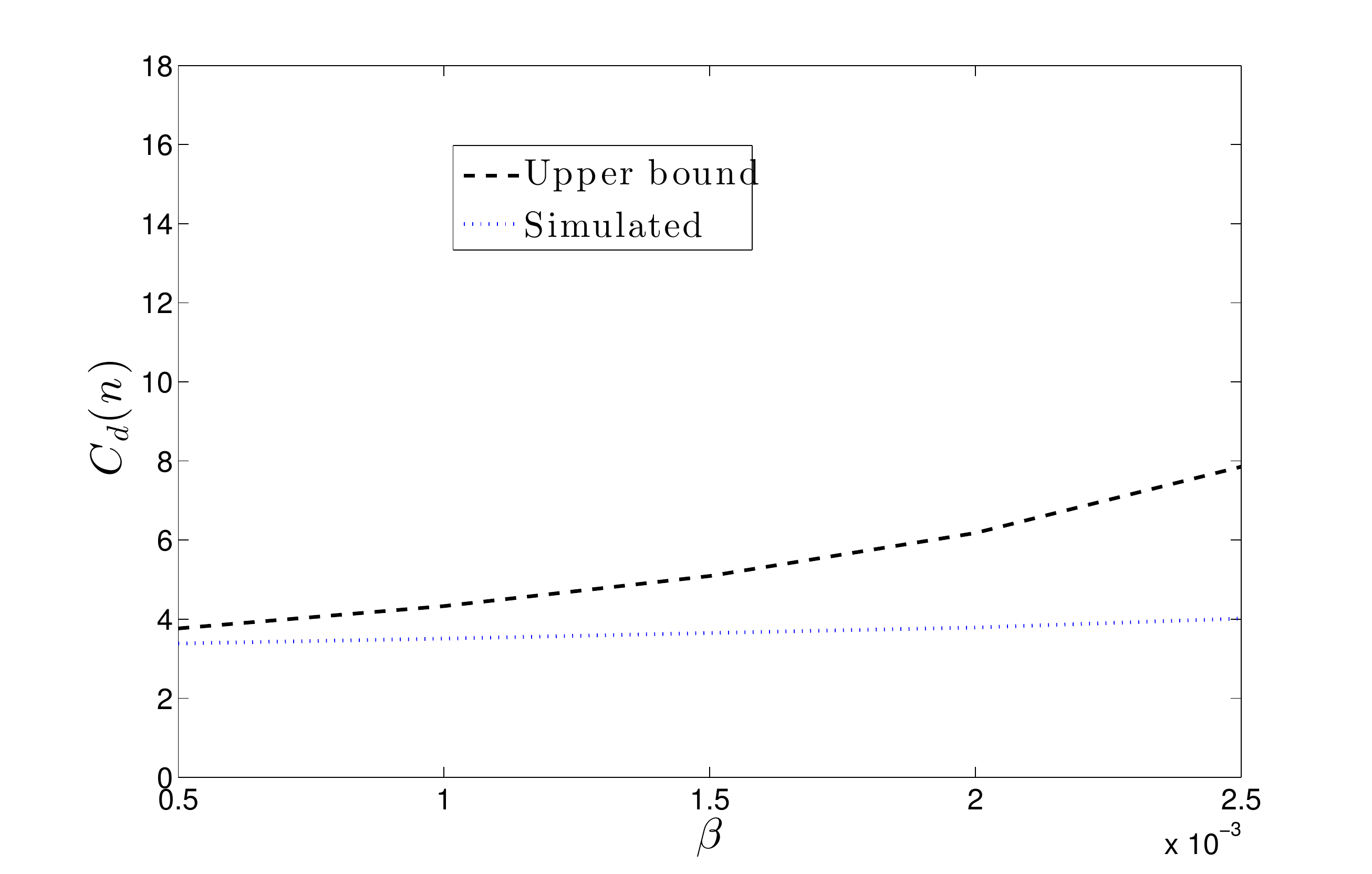}}
 \caption{Simulated disease cost on Caltech and Wikipedia networks}
  \label{fig:realSim}
\end{figure}

Simulating the disease cost on both of these networks, similarly to the method used for the random networks in the previous section, we see that our model and upper bound are relatively close to the simulated cost.  See Figure \ref{fig:realSim} for the performance on both the Caltech and Wikipedia social networks.  As we do not have access to a $p(\cdot)$ for either of these real networks, we focused on the difference between the upper bound (Theorem \ref{thm:bound}) and the actual simulation of the disease.  Since the Wikipedia network was rather large, we only plot the upper bound, using the maximum eigenvalue of the system matrix, for that network.

%% file: extensions.tex
\section{Extensions}
\label{sec:extensions}

One of the key benefits of the relatively simple expressions of the disease cost in Section \ref{sec:results} is the ability to calculate and minimize the total ``social cost'' of an epidemic, defined as the sum of the disease cost and the cost of whatever containment or immunization scheme is being considered.  As a first step, we examine a one-shot immunization scheme.  In this scenario, nodes are immunized  at $t=0$ and remain immune for all time, incurring a single immunization cost.  This cost could represent the monetary cost of a vaccine to an individual, the cost of quarantining, or a normalized development and administration cost.  For simplicity, we assume that these costs can all be represented by a single quantity, $c_v$.

Here, we incorporate the immunization process into the random graph model from Section \ref{sec:netmodel}.  Recall that to generate a network with $n$ nodes according to this model, the degree distribution $p_n(\cdot)$ is sampled $n$ times.  Consider a randomized immunization procedure where $\pi n$ nodes are chosen uniformly at random at $t=0$ to be immunized.  The resulting network is formed by sampling $p_n(\cdot)$ only for the nodes that remain after the immunization.  The immunized nodes are removed from the adjacency matrix $A$ and the corresponding system matrix $M$, yielding $\tilde{A}$ and $\tilde{M}$, respectively, with $\E[\dim \tilde{M}] = (1-\pi) n := \tilde{n}$.  To model a degree-based immunization scheme, we can simply truncate the degree distribution $p_n(\cdot)$ to generate the network as before.  In either case, the social cost of an epidemic can be defined as
\begin{equation} S(M,\tilde{M}) := \frac{1}{n} \left[(\dim M - \dim \tilde{M}) c_v + \left(1^T(I-\tilde{M})^{-1}1\right) \alpha c_d \right]. \end{equation}
The second half of the above expression represents the disease cost on the immunized network and can be calculated using the results in Section \ref{sec:results}. In general, any one-shot immunization scheme simply results in a transformation of $p_n(\cdot)$ and the disease cost calculations in Section \ref{sec:results} still apply.

\subsection{Optimal random immunization}
For the epidemic process $(G_{n, p_n(\cdot)}, \delta, \beta_n, \alpha, c_d)$, consider a one-shot random immunization scheme. Define the expected per node social cost $S(\pi)$ as a function of the fraction $\pi$ of immunized nodes as follows:
\begin{equation}
S(\pi) := \pi c_v + (1- \pi) C_D\left(n (1-\pi) \right)
\end{equation}
Applying the results from Section \ref{sec:results}, we can obtain results for $S (\pi)$. We illustrate this with an Erd\"{o}s-R\'{e}nyi random network. Using \eqref{eq:erExactCost} from Section \ref{sec:ER} for $G_{n,  \boldsymbol{\delta}(w-np)}$, we have that, as $n\to\infty$,
\begin{equation} \label{eq:socCostER}
S (\pi) = \pi c_v +  \frac{(1-\pi) \alpha c_d}{\delta - \beta n(1-\pi) p} \text{ w.h.p.}
\end{equation}
We can now determine the optimal fraction of nodes to immunize by minimizing \eqref{eq:socCostER}. For convenience, we normalize $c_d = 1$ and define $a := \frac{\alpha}{\delta}$, $b := \frac{\alpha\delta}{\left(\delta - \beta n p \right)^2}$ and $c := c_v/ c_d$. The optimal fraction of nodes to immunize is then
\begin{equation*}
\pi_{opt} =  \begin{cases} 1 & c \leq a < b \\
1 - \frac{\delta - \sqrt{\delta \alpha / c }}{\beta n p} & a < c < b \\
0 & a < b \leq c \end{cases} \end{equation*}

\begin{figure}[htp]
	\centering
  \subfloat[Small $c$]{\label{fig:ssC}\includegraphics[width=0.33\textwidth]{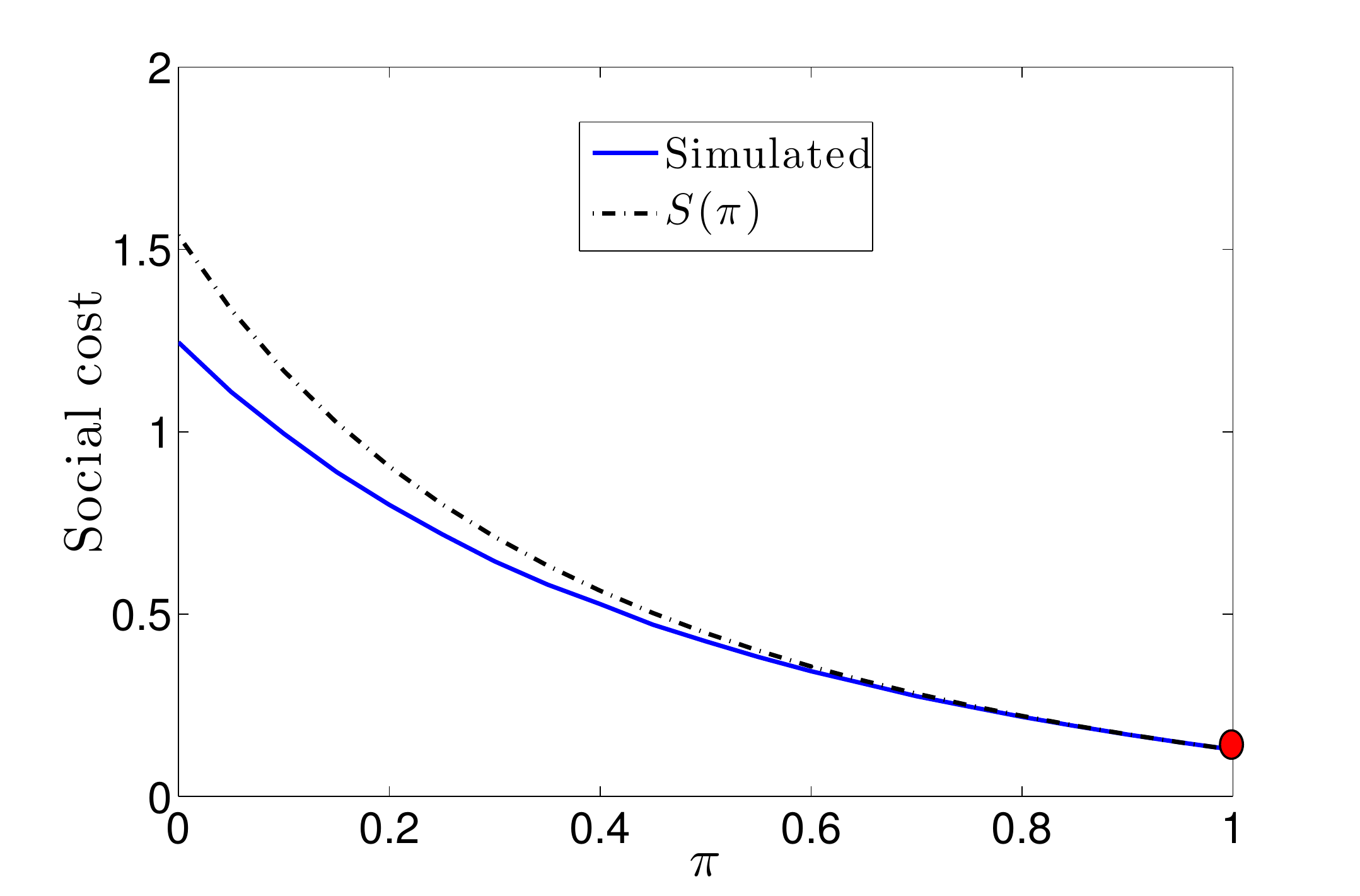}}
  \subfloat[Medium $c$]{\label{fig:smC}\includegraphics[width=0.33\textwidth]{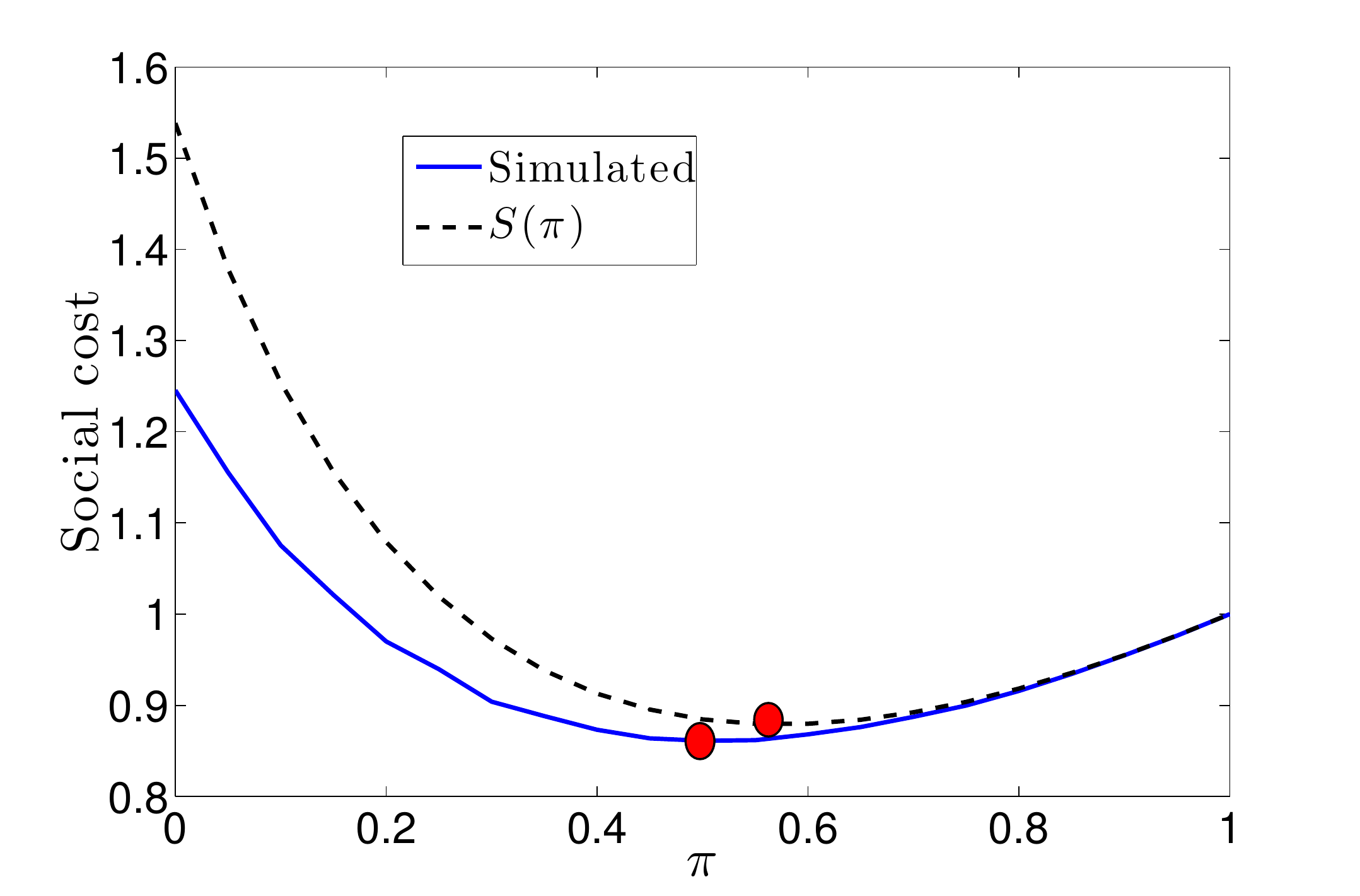}}
  \subfloat[Large $c$]{\label{fig:slC}\includegraphics[width=0.33\textwidth]{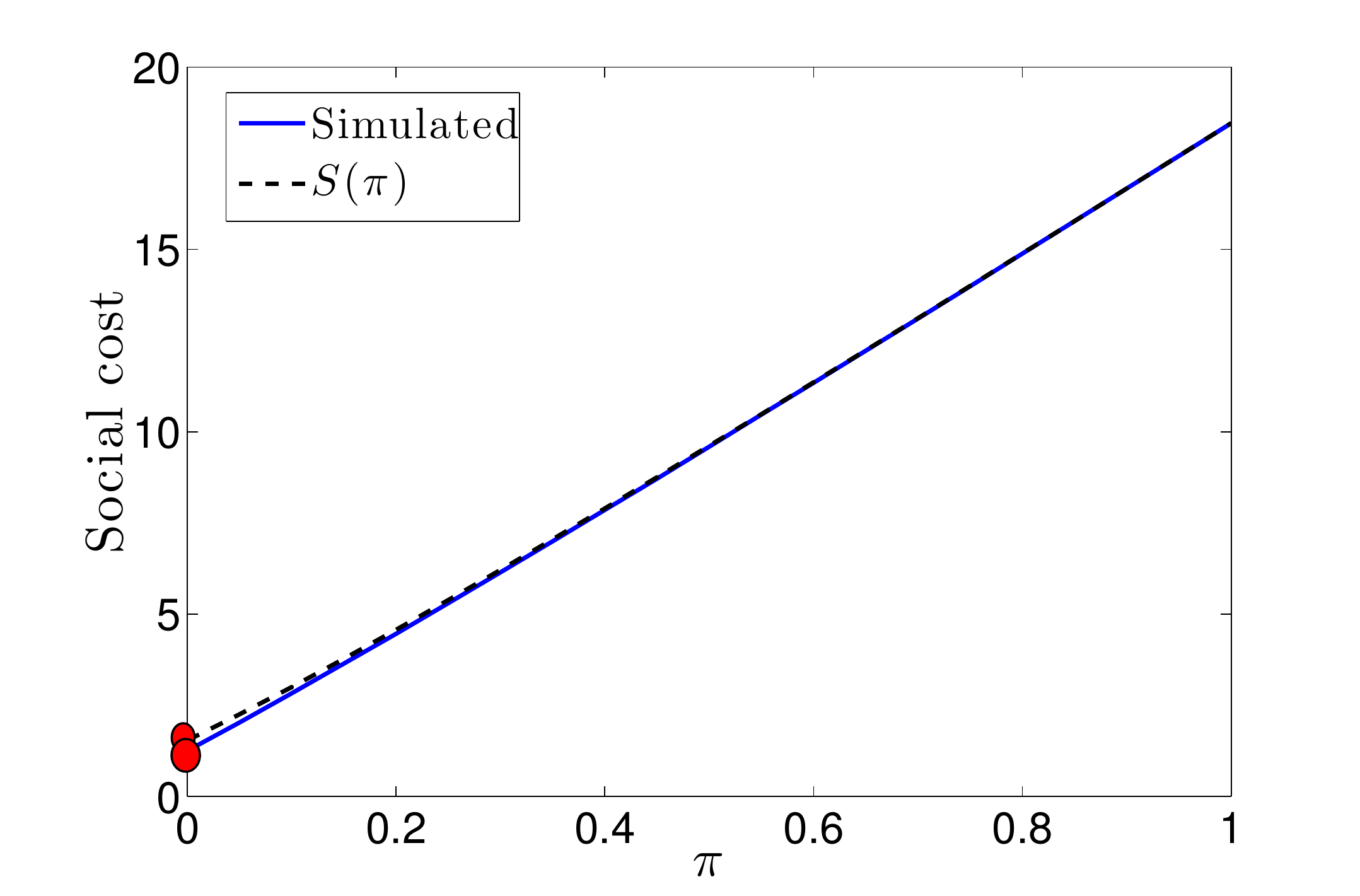}}
  \caption{Social cost simulations for different values of $c$ on Erd\"{o}s-R\'{e}nyi network as a function of the immunization probability $\pi$.  The optimal immunization probability in each case is highlighted with a red circle.}
  \label{fig:scSim}
\end{figure}

To illustrate the above, we simulate a disease propagating on an Erd\"{o}s-R\'{e}nyi graph, with $n = 10^5$, $p=1.27 \times 10^{-4}$, $\alpha = 0.20$, $\beta = 0.02$, and $\delta = 0.39$, according to the infection model described in Section \ref{sec:infmodel}.  We use a low $c = 0.13$, medium $c = 1.00$, and high $c = 18.46$. The simulated cost as a function of $\pi$ is shown in Figure \ref{fig:scSim}, together with the approximate calculated cost as given in \eqref{eq:socCostER}.

%% file: conclusion.tex
\section{Conclusion}
\label{sec:conc}

In this paper we have adapted tools from random matrix theory in order to quantify the economic impact of an epidemic on a complex network.  Using a linearized dynamical system based on the popular SIS model and a random graph as the underlying network, we have provided results that characterize the cost of the disease in the large graph limit and we have derived bounds for the disease cost for a given graph. This cost depends on the entire transient behavior of the system and hence this analysis differs from previous work that focuses on the steady-state equilibrium.   To highlight the usefulness of our results, we have included a brief application to the analysis of optimal immunization strategies.

Our analysis adapts ideas and techniques from random matrix theory to the study of epidemics, which differentiates this work from previous research in the area.  The analysis shows that random matrix theory can provide powerful tools for this domain, but that they must be adapted significantly in order to apply.  Thus, we hope that this paper motivates further research toward developing a strong connection between these domains.
  

%% file: appProofs.tex

\label{sec:appProofs}
In this section, we prove Lemma \ref{lemma:allTerms}. Recall the following relations.
\begin{align*}
	F &= \int_{v_{min}}^{v_{max}} \frac{p(v)}{v^{-1} - \kappa^2 F }dv, \\
	\kappa &= \lim_{n \to \infty} \sqrt{\frac{\beta_n}{\delta^2 \bar{v}}}, \\
	\bar{v} &= \lim_{n \rightarrow \infty} \frac{1}{n\mu} \as,
\end{align*}
We start by showing the following technical result. Let $Z_{n \times 1}$ be a vector with independent entries that satisfies $\E Z_k = 0$, $\E Z_k^2 = 1/n$ and the distribution of $n Z_k$ has bounded support. Then,
\begin{equation}
\label{eq:techClaim}
\lim_{n \to \infty} \left[ Z^T \left(V^{-1} - \sqrt{\frac{\beta_n}{\delta^2 \bar{v}}} C\right)^{-1} Z \right]  = \lim_{n \to \infty}  \frac{1}{n} \E \tr \left(  V^{-1} - \sqrt{\frac{\beta_n}{\delta^2 \bar{v}}} C \right)^{-1} = F \as,
\end{equation}
where $V$ is as defined in \eqref{eq:defV} and $C$ is the Wigner matrix defined in \eqref{eq:defC}. The argument is similar to the derivation of Wigner's Semicircle Law using the Stieltjes transform method in random matrix theory as in \citet{taoBook}. For convenience, define
$$ \kappa_n := \sqrt{\frac{\beta_n}{\delta^2 \bar{v}}}.$$
To prove \eqref{eq:techClaim}, we first characterize the behavior of $Z^T R Z$ for a deterministic positive semidefinite matrix $R$ with operator norm $\bigo (1)$, where $\bigo (\cdot)$ is the standard ``big-O'' notation as in \citet{horn2005matrix, taoBook}. In particular, we use the following concentration inequality; see \citet{talagrand} for its proof.
\begin{theorem_nonum}[Talagrand's Inequality:] Let $K > 0$, and let $Y_1,\ldots, Y_n$	be	 independent variables with $| Y_i | \leq K$ for all $1\leq i \leq n$. Let $F: \mathbb{R}^n \to \mathbb{R}$ be a 1-Lipschitz convex function. Then there exists positive constants $B, b$ such that for any $\mu$,
\begin{align*}
\Pr \Big[ \ \left\vert F(Y) - \M F(Y) \right\vert \geq  \mu \ \Big] \ \leq \ & B \exp(-b \mu^2/ K^2) \quad\text{a.s.},
\end{align*}
where $\mathbb{M}$ denotes the median.
\end{theorem_nonum}
Applying this result on the 1-Lipshitz function $ \| \sqrt{n} R^{1/2} Z \| $, there exists positive constants, $\mu, B, b$ for which, we have:
\begin{align}
\label{eq:talag} \Pr \Big[ \ \left\vert (Z^T R Z)^{1/2} - \M (Z^T R Z)^{1/2} \right\vert \geq \mu/ \sqrt{n} \ \Big] \leq B \exp(-b \mu^2) \quad\text{a.s.}
\end{align}
Now, almost surely $\| Z \| = 1$ and $\| R \| = 1$ and hence $(Z^T R Z)^{1/2} = \bigo(1)$ and $(Z^T R Z)^{1/2} = \bigo(1)$. The relation in \eqref{eq:talag} states that the difference between the two $\bigo(1)$ terms is $\bigo(1/\sqrt{n})$ almost surely, i.e.,
$$(Z^T R Z)^{1/2} = \M (Z^T R Z)^{1/2} + \bigo(1/ \sqrt{n}) \as$$
Squaring and rearranging, we have:
$$Z^T R Z = \M (Z^T R Z) + \bigo(1/ \sqrt{n}) \as$$
Replacing median with the expectation as in \citet{taoBook}, we have
\begin{align} \label{eq:int.1} Z^T R Z = \E (Z^T R Z) + \bigo(1/ \sqrt{n}) \as \end{align}
The above result holds for a symmetric positive semidefinite matrix $R$ with operator norm $\bigo(1)$. The result can be extended to symmetric sign indefinite matrices $R$ with operator norm $\bigo (1)$ as follows. Let $R = R_+ - R_-$, where $R_+$ and $R_-$ are positive semidefinite matrices each with operator norms $\bigo (1)$. (Note that $R_+$ and $R_-$ are uniquely determined by the eigen decomposition of $R$.) Applying the relation in \eqref{eq:int.1} to $R_+$ and $R_-$ and adding, we get that \eqref{eq:int.1} holds for any symmetric matrix $R$ with operator norm $\bigo(1)$. Also, since $\E Z = 0$, we have
$$ \E Z^T R Z = \E \sum_{i,j = 1}^n Z_{i}Z_{j} R_{ij} = \frac{1}{n}\sum_{i = 1}^n R_{ii} = \frac{1}{n}\tr (R). $$

For the purpose of our proof, we argue that the matrix $(V^{-1} - \kappa_n C)^{-1}$ has operator norm $\bigo(1)$ almost surely. The system matrix $M$ is stable and hence $I - M \succeq \epsilon I$ for some $\epsilon > 0$, where for two matrices $M_1$ and $M_2$, $M_1 \succeq M_2$ denotes $M_1 - M_2$ is positive semidefinite. From \eqref{eq:defX}, it follows that
$$ I - \sqrt{\frac{n \beta_n \mu}{\delta^2}} V^{1/2} C V^{1/2} \ \succeq \ \epsilon I + \mu v v^T \ \succeq \ \epsilon I. $$
Rearranging the above relation, we get
$$ V^{-1} - \sqrt{\frac{n \beta_n \mu}{\delta^2}} C \ \succeq \ \frac{\epsilon}{\delta} V^{-1} \ \succeq \ \frac{\epsilon v_{max}^{-1}}{\delta} I. $$
Also, $\sqrt{{n \beta_n \mu}/\delta^2}  -  \kappa_n \to 0$ almost surely as $n \to \infty$ and hence the eigendistribution of the matrix $(V^{-1} - \kappa_n C)$ is bounded away from zero for sufficiently large $n$. Also, we have
$$ V^{-1} - \kappa_n C \ \preceq v_{min}^{-1} I - \kappa_n C $$
Note that for large $n$, the Wigner matrix $C$ has a distribution close to the semi-circle law that has bounded support. Informally, the eigendistribution of the matrix $(V^{-1} - \kappa_n C)$ is bounded above and below by positive numbers and so is the case for its inverse. Thus, the operator norm of $(V^{-1} - \kappa_n C)^{-1}$ is $\bigo(1)$. To apply the relation \eqref{eq:int.1} to this matrix, note that $V$ and $C$ are independent of $Z$. Thus, conditioning on the matrix $(V^{-1} - \kappa_n C)^{-1}$, we have
\begin{align} \label{eq:int.2} Z^T (V^{-1} - \kappa_n C)^{-1} Z = \frac{1}{n} \tr (V^{-1} - \kappa_n C)^{-1} + \bigo(1/ \sqrt{n}) \as \end{align}

Note that the term $\tr (V^{-1} - \kappa_n C)^{-1}$ on the right hand side of the relation in \eqref{eq:int.2} is a random variable. Next, we relate this to its expected value using a concentration inequality. For a real variable $x$, define
$$ s_n (x) := \tr (V^{-1} - \kappa_n C)^{-1}.$$
The function $s_n (x)$ is a continuous and differentiable function of $x$ for $x = \kappa_n$ since the eigendistribution of $(V^{-1} - \kappa_n C)^{-1}$ is bounded above and below by positive constants. Let the eigenvalues of $(V^{-1} - \kappa_n C)$ be $\lambda_1 \leq \lambda_2 \leq \ldots \leq \lambda_n$. Also, let the eigenvalues of the principal $(n-1) \times (n-1)$ sub-matrix of $(V^{-1} - \kappa_n C)$ be $\lambda'_1 \leq \lambda'_2 \leq \ldots \leq \lambda'_{n-1}$. By Cauchy's interlacing theorem (\citet{horn2005matrix}), these two sets of eigenvalues interlace with each other, i.e.,
$$ \lambda_1 \leq \lambda'_1 \leq \lambda_2 \leq \lambda'_2 \leq \ldots \leq \lambda'_{n-1} \leq \lambda_n.$$
Thus, $s_n(x)$ and $s_{n-1}(x)$ can be related as follows:
$$ s_n(x) - s_{n-1}(x) = \sum_{i=1}^n \frac{1}{\lambda_i} - \sum_{i=1}^{n-1} \frac{1}{\lambda'_i}.$$
The eigenvalues lie in a bounded interval that is bounded away from zero. The above expression can be bounded by the total variation of the function $1/x$ in the same interval of length $\bigo(1)$. Thus, adding a row and column to the $(n-1) \times (n-1)$ submatrix affects $s_n(x)$ by order $\bigo(1)$ almost surely. Now, we use the following concentration result on the function $\frac{1}{n} s_n(x)$.
\begin{theorem_nonum}[MacDiarmid's Inequality:] Let $Y_1,\ldots, Y_n$ be independent random variables taking values from the set $\mathcal{Y}$. Further let $F: \mathcal{Y}^n \to \mathbb{R}$ be a function that satisfies
$$ \vert F(Y_1, \ldots, Y_{i-1}, Y_i, Y_{i+1}, \ldots, Y_n) -  F(Y_1, \ldots, Y_{i-1}, Y'_i, Y_{i+1}, \ldots, Y_n) \vert \leq b_i$$
for all $Y_1, \ldots, Y_n$, $Y'_i$ in $\mathcal{Y}$. Then
\begin{align*}
\Pr \Big[ \ \left\vert F(Y) - \E F(Y) \right\vert \geq  \mu \ \Big] \ \leq \ & \exp\left(-\frac{\mu^2} {\sum_{i=1}^n b_i^2}\right) \as.
\end{align*}
\end{theorem_nonum}
Applying this result on $\frac{1}{n} s_n(x)$, where the $Y_i$'s are the columns (or rows) of the symmetric matrix $(V^{-1} - \kappa_n C)^{-1}$, it can be checked that the following holds:
\begin{align} \label{eq:int.3} \frac{1}{n} \tr (V^{-1} - \kappa_n C)^{-1} = \E \frac{1}{n} \tr (V^{-1} - \kappa_n C)^{-1} + \bigo (1/\sqrt{n}) \as. \end{align}
Combining \eqref{eq:int.2} and \eqref{eq:int.3}, we have 
\begin{align} \label{eq:int.4}  Z^T (V^{-1} - \kappa_n C)^{-1} Z = \frac{1}{n} \E \tr (V^{-1} - \kappa_n C)^{-1}  + \bigo(1/\sqrt{n}) \as \end{align}

Next, we characterize $\frac{1}{n} \E \tr \left( V^{-1} - \kappa_n C\right)^{-1}$ as $n \to \infty$. For convenience, define the function $T_n(x)$ for the real variable $x$ as follows.
\begin{align}
\label{eq:defT}
T_n (x) := \frac{1}{n} \E \tr \left( V^{-1} - x C\right)^{-1}.
\end{align}
We start with the block matrix decomposition of the $n \times n$ matrices $V$ and $C$. Let
\begin{align}
\label{eq:defCW}
V =\begin{bmatrix}
       v_1 & 0 \\
       0 & V_2
     \end{bmatrix}
\quad \text{and} \quad
C =\begin{bmatrix}
       c_{11} & C_{21}^T \\
       C_{21} & C_{22}
     \end{bmatrix},
\end{align}
where $v_1$ and $c_{11}$ are scalars and the rest are matrices of appropriate sizes. From the Matrix Inversion Lemma (\citet{horn2005matrix}), it follows that
\begin{align}
T_n (\kappa_n)
& =   \frac{1}{n} \E \tr
	{\begin{bmatrix}
       		v_1^{-1} - \kappa_n c_{11} & - \kappa_n C_{21}^T \\
       		- \kappa_n C_{21} & \underbrace{ V_2^{-1} - \kappa_n C_{22}}_{:=D}
     	\end{bmatrix}}^{-1}  \notag\\
& =   \frac{1}{n} \E \tr
	\begin{bmatrix} \label{inv_matrix2}
		\Delta^{-1} &	-\Delta^{-1}C_{21}^{T} D^{-1} \\
		- D^{-1} C_{21} \Delta^{-1}  & D^{-1} + D^{-1} C_{21} \Delta^{-1} C_{21}^T D^{-1}
	\end{bmatrix}.
\end{align}
where $\Delta$ is the Schur complement \citet{horn2005matrix} of $D$ defined as
\begin{align}
	\Delta :=  v_1^{-1} - \kappa_n c_{11}  - \kappa_n^2 C_{21}^T D^{-1} C_{21}.
\end{align}
The diagonal entries of the matrix $\left( V^{-1} - \kappa_n C \right)^{-1}$ are identically distributed and thus we have
\begin{align}
T_n (\kappa_n)
& = \E \Delta^{-1} \notag\\
& = \E \left[ v_1^{-1} - \kappa_n c_{11}  - \kappa_n^2 C_{21}^T D^{-1} C_{21} \right] ^{-1}  \notag\\
& = \E \left[ v_1^{-1} - C_{21}^T \left( V_2^{-1} -\kappa_n C_{22}  \right)^{-1} C_{21} +  \bigo \left(1/n\right ) \right] ^{-1} \label{eq:Tn}.
\end{align}
since $c_{11}$ has mean zero and variance $1/{n}$. Note that $C_{22}\sqrt{\frac{n}{n-1}}$ is a Wigner matrix of size $(n-1) \times (n-1)$. Using $Z := C_{21}\sqrt{\frac{n}{n-1}}$ in \eqref{eq:int.4}, we get
\begin{align}
\label{eq:C21}
C_{21}^T \left(  V_{2}^{-1} - \kappa_n C_{22} \right)^{-1} C_{21}
& = \left( \frac{n}{n-1} \right)^ {3/2} T_{n-1} \left( \kappa_n \sqrt{\frac{n}{n-1}} \right) + \bigo\left( \frac{1}{\sqrt{n}}\right) \quad\text{a.s.}
\end{align}
Informally, $T_n(\kappa_n)$ and $T_{n-1}(\kappa_n)$ differ by $\bigo(1/\sqrt{n})$ almost surely. This follows from the fact that $T_n(x) = \frac{1}{n} \E s_n(x)$ and we have already established that $s_n(x)$ and $s_{n-1}(x)$ differ by $\bigo(1)$ almost surely. Also, it can be checked that $T_n(x)$ is continuous and differentiable and hence we have
\begin{align*}
T_{n-1} \left( \kappa_n \sqrt{\frac{n}{n-1}} \right)
&= T_{n} \left( \kappa_n \sqrt{\frac{n}{n-1}} \right)  + \bigo(1/\sqrt{n}) \as \\
& = T_{n} \left( \kappa_n \right) +  \bigo(1/\sqrt{n}) \as
\end{align*}
Using the above equality in \eqref{eq:C21}, we have
\begin{align*}
C_{21}^T \left(  V_{2}^{-1} - \kappa_n C_{22} \right)^{-1} C_{21}
 = T_{n} \left( \kappa_n \right) + \bigo\left( \frac{1}{\sqrt{n}}\right) \quad\text{a.s.}.
\end{align*}
Replacing the expression for $C_{21}^T \left(  V_{2}^{-1} - \kappa_n C_{22} \right)^{-1} C_{21}$ in \eqref{eq:Tn}, we have the following implicit relation for $T_n (\kappa_n)$:
\begin{align*}
T_n(\kappa_n)
& = \E \left[\frac{1}{v_{1}^{-1} - \kappa_n^2 T_{n} (\kappa_n) } \right] + \bigo\left(\frac{1}{\sqrt{n}} \right)  \as
\end{align*}
Note that the expectation is over $v_1$ that is drawn according to the distribution $p(\cdot)$. Taking the limit as $n \to \infty$ and appealing to the Borel-Cantelli lemma as in \citet{taoBook} finishes the proof for \eqref{eq:techClaim}.




Next, we compute the terms $\frac{1}{n} ( 1^T X^{-1} v)$, $\frac{1}{n} (1^T X^{-1} 1 )$, and $\frac{1}{n} (v^T X^{-1} v)$ in terms of $\delta$, $\kappa$ and $p(v)$. Assumption \ref{ass:expTrace} holds throughout. Recall the following relation:
\begin{align*}
	X & = \delta I - \sqrt{n \beta_n \mu} V^{1/2} C V^{1/2}.
\end{align*}

\subsection*{Proof of Lemma \ref{lemma:allTerms} (a):}
\noindent First we prove
\begin{equation} \label{eq:claim1}
	\lim_{n \to \infty} \frac{1}{n} 1^T X^{-1} v  = \frac{F}{\delta} \quad\text{a.s.},
\end{equation}

\begin{align}
\lim_{n \to \infty} \frac{1}{n} 1^T X^{-1} v
& = \lim_{n \to \infty} \frac{1}{n} [1^{T} X^{-1} V 1]  \notag\\
& = \lim_{n \to \infty}\frac{1}{n} \left[ 1^{T} ( \delta I - \sqrt{n \beta_n \mu} V^{1/2} C V^{1/2} )^{-1} V 1 \right] \notag\\
& =  \lim_{n \to \infty} \frac{1}{n} \left[ 1^{T} ( \delta V^{-1} - \sqrt{n \beta_n \mu} V^{-1/2} C V^{1/2} )^{-1} 1 \right]  \notag\\
& = \lim_{n \to \infty}  \frac{1}{n} \left[ 1^T( \delta V^{-1} - \sqrt{n \beta_n \mu} C)^{-1} 1 \right] \notag\\
& = \frac{1}{\delta} \lim_{n \to \infty}  \frac{1}{n} \left[ 1^T \left( V^{-1} - \sqrt{\frac{\beta_n}{\delta^2 \bar{v}}} C \right)^{-1} 1 \right] \notag\\
& = \frac{1}{\delta} \lim_{n \to \infty}  \frac{1}{n} \E \tr \left(V^{-1} - \sqrt{\frac{\beta_n}{\delta^2 \bar{v}}} C \right)^{-1} \as \label{eq:useAss.1} \\
& = \frac{F}{\delta} \as \label{eq:useLemma.1}
\end{align}
The equality in \eqref{eq:useAss.1} follows from Assumption \ref{ass:expTrace} and the one in \eqref{eq:useLemma.1} follows from \eqref{eq:techClaim}. This completes the proof of Lemma \ref{lemma:allTerms} (a).

\subsection*{Proof of Lemma \ref{lemma:allTerms} (b):}
Next, we compute the term $\frac{1}{n} 1^T X^{-1} 1$ and show that
\begin{equation}
\label{eq:claim2} \lim_{n \to \infty} \frac{1}{n} 1^T X^{-1} 1 = \frac{1 + \kappa^2 F^2 }{ \delta} \quad \text{a.s.}
\end{equation}
Substituting the expression for $X$, we have
\begin{align*}
\lim_{n \to \infty} \frac{1}{n} 1^T X^{-1} 1
& = \lim_{n \to \infty} \frac{1}{n} \left[ 1^{T} \left( \delta I -  \sqrt{n \beta_n \mu} V^{1/2} C V^{1/2} \right)^{-1} 1 \right] \\
& = \frac{1}{\delta} \lim_{n \to \infty} \frac{1}{n} \left[ 1^{T} \left( I - \sqrt{\frac{\beta_n}{\delta^2 \bar{v}}} V^{1/2} C V^{1/2} \right)^{-1} 1 \right] \quad\text{a.s.}\\
& = \frac{1}{\delta} \lim_{n \to \infty} \frac{1}{n} \E \tr \left( I - \sqrt{\frac{\beta_n}{\delta^2 \bar{v}}} V^{1/2} C V^{1/2} \right)^{-1} \quad\text{a.s.}
\end{align*}
where the last step follows from Assumption \ref{ass:expTrace}. Now, we use the same block matrix decomposition of $C$ and $V$ as above. The definitions are restated here for convenience.
\begin{align*}
V =\begin{bmatrix}
	v_1 & 0 \\
       	0 & V_2
     \end{bmatrix}
\quad \text{and} \quad
C =\begin{bmatrix}
       c_{11} & C_{21}^T \\
       C_{21} & C_{22}
     \end{bmatrix},
\end{align*}
where $v_1$ and $c_{11}$ are scalars and the rest are matrices of appropriate sizes.  Writing the matrix using these expressions, we have
\begin{align*}
\left( I - \sqrt{\frac{\beta_n}{\delta^2 \bar{v}}}  V^{1/2} C V^{1/2} \right)^{-1}
& = {\begin{bmatrix}
       1 - \sqrt{\frac{\beta_n}{\delta^2 \bar{v}}} v_1 c_{11} & -\sqrt{\frac{\beta_n}{\delta^2 \bar{v}}} \sqrt{v_1} C_{21}^T V_2^{1/2}\\
       -\underbrace{\sqrt{\frac{\beta_n}{\delta^2 \bar{v}}} \sqrt{v_1} V_2^{1/2} C_{21}}_{:=y}  & \underbrace{I - \sqrt{\frac{\beta_n}{\delta^2 \bar{v}}}  V_2^{1/2} C_{22} V_2^{1/2}}_{:=\tilde{D}}
     \end{bmatrix}}^{-1}.
\end{align*}
\noindent Applying the Matrix Inversion Lemma and continuing, we have
\begin{align}
\lim_{n \to \infty} \E \tr \frac{1}{n} & \left( I - \sqrt{\frac{\beta_n}{\delta^2 \bar{v}}}  V^{1/2} C V^{1/2} \right)^{-1} \notag\\
& = \lim_{n \to \infty} \E \left[ 1 -  \sqrt{\frac{\beta_n}{\delta^2 \bar{v}}}  v_1 c_{11} -  y^T \tilde{D}^{-1} y   \right]^{-1} \notag\\
& = \lim_{n \to \infty} \E \left[ 1 -  \sqrt{\frac{\beta_n}{\delta^2 \bar{v}}}  v_1 c_{11}  - \frac{\beta_n}{\delta^2 \bar{v}} v_1 C_{21}^T  \left(V_2^{-1} - \sqrt{\frac{\beta_n}{\delta^2 \bar{v}}} C_{22} \right)^{-1} C_{21}   \right]^{-1} \notag\\
& = \lim_{n \to \infty} \E \left[ 1 - \frac{\beta_n}{\delta^2 \bar{v}} v_1 C_{21}^T  \left(V_2^{-1} - \sqrt{\frac{\beta_n}{\delta^2 \bar{v}}} C_{22} \right)^{-1} C_{21}   \right]^{-1} \notag\\
& = \E \left( 1- \kappa^2 v_1 F  \right)^{-1} \quad \text{a.s.} \label{applyB1_1} \\
& = \int_{v_{min}}^{{v_{max}}} \frac{v^{-1}p(v)}{ v^{-1} - \kappa^2 F} dv \quad \text{a.s.}.  \notag
\end{align}
where \eqref{applyB1_1} follows from \eqref{eq:techClaim}. For convenience, define
\begin{equation} \label{eq:S1_def}
	S_1 := \int_{v_{min}}^{{v_{max}}} \frac{v^{-1}p(v)}{ v^{-1} - \kappa^2 F} dv.
\end{equation}

\noindent Consider the following relation:
\begin{align*}
S_1 - \kappa^2 F^2
& =\int_{v_{min}}^{{v_{max}}} \frac{v^{-1}p(v)}{ v^{-1} - \kappa^2 F} dv    - \kappa^2 \int_{v_{min}}^{{v_{max}}} \frac{F p(v)}{ v^{-1} - \kappa^2 F} dv\\
& = \int_{v_{min}}^{{v_{max}}} p(v) dv\\
& = 1.
\end{align*}
Rearranging and solving for $S_1$ in terms of $F$, we see that $S_1 = 1 + \kappa^2 F^2$ and
\begin{equation*}
	\lim_{n \to \infty} \frac{1}{n} 1^T X^{-1} 1  =  \frac{S_1}{\delta} = \frac{1 + \kappa^2 F^2 }{ \delta} \quad \text{a.s.},
\end{equation*}
This proves Lemma \ref{lemma:allTerms} (b).

\subsection*{Proof of Lemma \ref{lemma:allTerms} (c):}
Now we move onto the last term, i.e., we prove that
\begin{equation} \label{eq:claim3}\lim_{n \to \infty} \frac{1}{n} v^T X^{-1} v = \begin{cases}
	\E v^2 \quad\text{a.s.} & \text{if } \kappa = 0,\\
	\frac{1}{\delta \kappa^2} \left( 1- \frac{\bar{v}}{F} \right) \quad\text{a.s.} & \text{if } \kappa \neq 0.
	\end{cases}\end{equation}
For brevity, we outline the steps involved.

\begin{align*}
\lim_{n \to \infty} \frac{1}{n} v^T X^{-1} v
& =  \lim_{n \to \infty} \frac{1}{n} \left[ 1^{T} V \left( \delta I -  \sqrt{n \beta_n \mu} V^{1/2} C V^{1/2}  \right)^{-1} V 1 \right] \\
& = \frac{1}{\delta} \lim_{n \to \infty} \frac{1}{n} \left[ 1^{T} V^{1/2} \left( V^{-1} - \sqrt{\frac{\beta_n}{\delta^2 \bar{v}}} C \right)^{-1} V^{1/2} 1 \right] \\
& = \frac{1}{\delta} \lim_{n \to \infty} \frac{1}{n} \E \tr \left[ V^{1/2} \left( V^{-1} - \sqrt{\frac{\beta_n}{\delta^2 \bar{v}}} C \right)^{-1} V^{1/2} \right] \as
\end{align*}
where the last step follows from Assumption \ref{ass:expTrace}. Using the block-decomposition of $C$ and $V$ as above,  and proceeding as before to expand the above expression, we get
\begin{align}
 \lim_{n \to \infty} \frac{1}{n} \E \tr & \left[ V^{1/2} \left( V^{-1} - \sqrt{\frac{\beta_n}{\delta^2 \bar{v}}} C \right)^{-1} V^{1/2} \right] \notag \\
& = \lim_{n \to \infty} \E v_1 \left[ v_1^{-1} -  \sqrt{\frac{\beta_n}{\delta^2 \bar{v}}} c_{11}  - \frac{\beta_n}{\delta^2 \bar{v}}  C_{21}^T  \left(V_2^{-1} - \sqrt{\frac{\beta_n}{\delta^2 \bar{v}}} C_{22} \right)^{-1} C_{21}   \right]^{-1} \notag \\
& = \lim_{n \to \infty} \E v_1 \left[ v_1^{-1} - \frac{\beta_n}{\delta^2 \bar{v}} C_{21}^T  \left(V_2^{-1} - \sqrt{\frac{\beta_n}{\delta^2 \bar{v}}} C_{22} \right)^{-1} C_{21}   \right]^{-1}  \notag\\
& = \E \left[ v_1 \left( v_1^{-1} - \kappa^2 F \right)^{-1} \right] \quad \text{a.s.} \label{applyB1_2} \\
& =  \int_{v_{min}}^{{v_{max}}} \frac{v p(v)}{ v^{-1} - \kappa^2 F} dv \quad \text{a.s.}. \notag
\end{align}
where \eqref{applyB1_2} follows from \eqref{eq:techClaim}. For convenience, define
\begin{equation} \label{eq:S2_def}
	S_2 := \int_{v_{min}}^{{v_{max}}} \frac{v p(v)}{ v^{-1} - \kappa^2 F} dv.
\end{equation}
\noindent If $\kappa = 0$, then
\begin{equation*} S_2  = \int_{v_{min}}^{{v_{max}}} v^2 p(v) dv  = \E v^2. \end{equation*}
\noindent For the case where $\kappa \neq 0$, consider the following relation:
\begin{align*}
 F - \kappa^2 S_2 F
& = \int_{v_{min}}^{{v_{max}}} \frac{p(v)}{ v^{-1} - \kappa^2 F} dv - \int_{v_{min}}^{{v_{max}}} \frac{\kappa^2  v Fp(v)}{ v^{-1} - \kappa^2 F} dv\\
& = \int_{v_{min}}^{{v_{max}}} vp(v) dw\\
& = \bar{v}.
\end{align*}
Rearranging and solving for $S_2$ in terms of $F$, we finally have
\begin{align*}
	\lim_{n \to \infty} \frac{1}{n} v^T X^{-1} v
	& =  \begin{cases}
	\E v^2 & \text{if } \kappa = 0,\\
	\frac{1}{\delta \kappa^2} \left( 1- \frac{\bar{v}}{F} \right) & \text{if } \kappa \neq 0.
	\end{cases}
\end{align*}
This proves Lemma \ref{lemma:allTerms} (c).


%
%
%
%